\documentclass[12pt,preprint]{aastex}
\newcommand{\ncm}{cm$^{-3}$}

\newcommand{\Ncm}{cm$^{-2}$}
\newcommand{\kms}{km s$^{-1}$}
\newcommand{\um}{$\mu$m}

\newcommand{\astH}[1]{H {\small #1}}

\newcommand{\Htwo}{H$_2$}
\newcommand{\HtwolineK}{H$_2$ $\upsilon=1\rightarrow0$ $S$(1)}

\newcommand{\astC}[1]{C {\small #1}}

\newcommand{\astNe}[1]{Ne {\small #1}}

\newcommand{\Rv}{$R_V$}

\newcommand{\nHone}{$n(\textrm{H I})$}
\newcommand{\nHtwo}{$n(\textrm{H}_2)$}

\newcommand{\nHe}{$n(\textrm{He})$}

\newcommand{\NHtwo}{$N(\textrm{H}_2)$}

\newcommand{\defNH}{$N_{\textrm{\tiny{H}}}$=$N$(\astH{I})+2$N$(\Htwo)}
\newcommand{\EvJ}{$E(\upsilon,J)$}

\newcommand{\vJ}{$\upsilon,J$}

\newcommand{\akari}{\textit{AKARI}}

\shorttitle{\Htwo{} Infrared Emission toward SNRs IC 443 and HB 21}
\shortauthors{Shinn et al.}

\begin{document}

\title{Ortho-to-Para Ratio Studies of Shocked \Htwo{} Gas in the Two Supernova Remnants IC 443 and HB 21}
%\title{Shocked \Htwo{} Gas with Non-equilibrium Ortho-to-Para Ratios Observed in the Two Supernova Remnants IC 443 and HB 21}

%\author{Jong-Ho Shinn et al.}
%\email{jhshinn@kasi.re.kr}
%\author{Jong-Ho Shinn\altaffilmark{1}, Bon-Chul Koo\altaffilmark{2}, Ho-Gyu Lee\altaffilmark{3}, Dae-Sik Moon\altaffilmark{3}}
\author{Jong-Ho Shinn\altaffilmark{1}, Ho-Gyu Lee\altaffilmark{2}, Dae-Sik Moon\altaffilmark{3}}

\email{jhshinn@kasi.re.kr, hglee@astron.s.u-tokyo.ac.jp, moon@astro.utoronto.ca}
%\email{jhshinn@kasi.re.kr}
\altaffiltext{1}{Korea Astronomy and Space Science Institute, 776 Daedeok-daero, Yuseong-gu, Daejeon, 305-348, Republic of Korea}
%\altaffiltext{2}{Dept. of Physics and Astronomy, FPRD, Seoul National University, 599 Gwanangno, Gwanak-gu, Seoul, 151-747, Republic of Korea}
\altaffiltext{2}{Dept. of Astronomy, Graduate School of Science, the University of Tokyo, 7-3-1 Hongo, Bunkyo-ku, Tokyo 113-0003, Japan}
\altaffiltext{3}{Dept. of Astronomy and Astrophysics, University of Toronto, Toronto, ON M5S 3H4, Canada}
%\altaffiltext{*}{Based on observations with \akari, a JAXA project with the participation of ESA.}

\begin{abstract}
We present near-infrared ($2.5-5.0$ \um) spectral studies of shocked \Htwo{} gas in the two supernova remnants IC 443 and HB 21, which are well known for their interactions with nearby molecular clouds.
The observations were performed with Infrared Camera (IRC) aboard the \akari{} satellite.
At the energy range 7000 K $\la$ \EvJ{} $\la$ 20000 K, the shocked H2 gas in IC 443 shows an ortho-to-para ratio (OPR) of $2.4^{+0.3}_{-0.2}$, which is significantly lower than the equilibrium value 3, suggesting the existence of non-equilibrium OPR. 
The shocked gas in HB 21 also indicates a potential non-equilibrium OPR in the range of $1.8-2.0$.
%The shocked \Htwo{} gas show level populations with probable non-equilibrium ortho-to-para ratios (OPRs) at the energy range 7000 K $\la$ \EvJ{} $\la$ 20000 K, which has never been reported before.
%The OPR for IC 443 is $2.4^{+0.3}_{-0.2}$, while the OPR for HB 21 is $\sim1.8-2.0$ but tentative.
The level populations are well described by the power-law thermal admixture model with a single OPR, where the temperature integration range is $1000-4000$ K.
We conclude that the obtained non-equilibrium OPR probably originates from the reformed \Htwo{} gas of dissociative J-shocks, considering several factors such as the shock combination requirement, the line ratios, and the possibility that \Htwo{} gas can form on grains with a non-equilibrium OPR.
We also investigate C-shocks and partially-dissociative J-shocks for the origin of the non-equilibrium OPR.
However, we find that they are incompatible with the observed ionic emission lines for which  dissociative J-shocks are required to explain.
The difference in the collision energy of H atoms on grain surfaces would make the observed difference between the OPRs of IC 443 and HB 21, if dissociative J-shocks are responsible for the \Htwo{} emission.
Our study suggests that dissociative J-shocks can make shocked \Htwo{} gas with a non-equilibrium OPR.
\end{abstract}


\section{Introduction \label{intro}}
Infrared \Htwo{} line emission is a useful tool for studying shock-cloud interactions.
\Htwo{} is one of basic coolants at the post-shock region \citep[cf.][]{Neufeld(1989)ApJ_340_869,Hollenbach(1989)ApJ_342_306,Kaufman(1996)ApJ_456_611}, and its infrared line emission usually suffers relatively small extinction effects \citep[cf.][]{Draine(2003)ARA&A_41_241}.
The shock-excited \Htwo{} line emission has been widely observed from diverse targets, such as protostellar jets and outflows, supernova remnants, and interacting galaxies (\citealt{Habart(2005)SSRv_119_71,Omont(2007)RPPh_70_1099}; and references therein).
Ground telescopes usually observe high excitation emission lines (\EvJ{} $\ga7000$ K) in the near-infrared wave bands, while space satellites observe low excitation emission lines (\EvJ{} $\la7000$ K) in mid-infrared.
The high excitation lines show the excitation temperature ($T_{ex}$) of a few thousand kelvin, while the low excitation lines show that of a few hundreds kelvin \citep[e.g.,][]{Rosenthal(2000)A&A_356_705,Rho(2001)ApJ_547_885,Giannini(2006)A&A_459_821,CarattioGaratti(2008)A&A_485_137,Shinn(2011)ApJ_732_124}.

\Htwo{} line emission manifests one important diagnostic parameter of \Htwo{} gas: the ortho-to-para ratio (OPR).
First we give a brief description of ortho- and para-\Htwo.
An \Htwo{} molecule consists of two identical proton nuclei, i.e., two identical fermions; hence, for a given nuclear spin combination, a specific rotational quantum number (space wave-function) is only permitted according to the Pauli exclusion principle \citep[cf.][]{Field(1966)ARA&A_4_207,Shull(1982)ARA&A_20_163}.
If the nuclear spins are summed to zero ($I=0$, singlet), it is called as para-\Htwo{} and only even rotational quantum number ($J$) is allowed.
On the other hand, if the nuclear spins are summed to one ($I=1$, triplet), it is called as ortho-\Htwo{} and only odd-$J$ is allowed.
The statistical weight ($g_J$) for a level (\vJ) is a multiplication of the nuclear spin degeneracy and the rotational degeneracy, $g_J=(2I+1)(2J+1)$; i.e., $g_J=3(2J+1)$ for ortho-\Htwo{} and $g_J=(2J+1)$ for para-\Htwo.

The OPR is the ratio between the abundance of ortho- and para-\Htwo{} in \emph{all} \vJ{} levels.
In general, lower \EvJ{} levels are much more populated than higher \EvJ{} levels \citep[cf.][]{Timmermann(1996)A&A_315_L281,Rosenthal(2000)A&A_356_705}.
Therefore, the OPR is mainly determined by the OPR at low \EvJ.
This means that the OPR can be different from the abundance ratio of ortho- and para-\Htwo{} in high \EvJ.
In this paper, we make a note on the energy level \EvJ{} when we mention the OPR at a specific \EvJ{} range; otherwise, the OPR means the abundance ratio of ortho- and para-\Htwo{} summed over all \vJ.
We also note that the OPR of \Htwo{} in local thermodynamic equilibrium (LTE) is equal to 3 at $T\ga300$ K, decreasing from 3 at $T\la300$ K \citep[cf.][]{Burton(1992)ApJ_399_563,Sternberg(1999)ApJ_516_371}.
When the OPR is different from this LTE OPR, we call it as a non-equilibrium OPR.

The OPR of \Htwo{} is useful in estimating thermal history of the gas \citep{Neufeld(2006)ApJ_649_816}.
Under typical conditions of molecular clouds, the collisional interconversion between ortho- and para-\Htwo{} is much slower than the collisional transitions between the same-type \Htwo{} (ortho-to-ortho and para-to-para).
Therefore, the OPR have the information on how much interconversion collisions the \Htwo{} gas experiences and what the initial OPR of the \Htwo{} gas is.
For shocked \Htwo{} gas, the possibility of the non-equilibrium OPR was briefly discussed by \cite{Chang(1991)ApJ_378_202}, and later this question was seriously addressed by \cite{Timmermann(1998)ApJ_498_246} and \cite{Wilgenbus(2000)A&A_356_1010}.
The last two studies predicted the existence of \Htwo{} gas with the non-equilibrium OPR in C-shocks or partially-dissociative J-shocks \citep[for the types of shocks, see][]{Draine(1993)ARA&A_31_373}.

Hitherto observations have revealed shocked \Htwo{} gas of non-equilibrium OPRs from mid-infrared observations, which trace low energy levels of \EvJ$\la7000$ K \citep[e.g.,][]{Neufeld(1998)ApJ_506_L75,Rodriguez-Fernandez(2000)A&A_356_695,Lefloch(2003)ApJ_590_L41,Neufeld(2006)ApJ_649_816,Quanz(2007)ApJ_658_487,Neufeld(2007)ApJ_664_890,Roussel(2007)ApJ_669_959,Hewitt(2009)ApJ_694_1266,Maret(2009)ApJ_698_1244,Neufeld(2009)ApJ_706_170}.
As far as we are aware, shocked \Htwo{} gas of non-equilibrium OPRs has never been observed at high energy levels of \EvJ$\ga7000$ K \citep[e.g.,][]{Gredel(1994)A&A_292_580,Smith(1997)A&A_327_1206,Giannini(2006)A&A_459_821,CarattioGaratti(2006)A&A_449_1077}.
Here we report two cases suggesting the shocked \Htwo{} gas with non-equilibrium OPRs at such high levels, which is revealed from near-infrared spectral observations with the \akari{} satellite.
%Here we report the shocked \Htwo{} gas of non-equilibrium OPRs at such high levels, which is revealed from near-infrared spectral observations with the \akari{} satellite.
The targets are two supernova remnants IC 443 (G189.1+3.0) and HB 21 (G89.0+4.7), which are well-known for their interactions with nearby molecular clouds (for more about the targets, see \citealt{Shinn(2009)ApJ_693_1883,Shinn(2010)AdSpR_45_445,Shinn(2011)ApJ_732_124}).
We present new HB 21 data here, while we used the previously presented data for IC 443 (clump B; i.e., IC 443B), which had been left for future analyses \citep{Shinn(2011)ApJ_732_124}.
The results are discussed with respect to different shock types, and we conclude that the obtained non-equilibrium OPR may originate from the reformed \Htwo{} gas of dissociative J-shocks.

\section{Observations and Data Reduction \label{obs-red}} 
The \akari{} near-infrared spectroscopic observations were performed on two regions of HB 21, where the shock-cloud interactions are under way.
The two regions, ``HB 21N'' and ``HB 21S'', are listed in Table \ref{tbl-obs}.
The ``HB 21N'' and ``HB 21S'' correspond to the regions ``N2front'' \citep{Shinn(2009)ApJ_693_1883} and ``S1'' \citep{Shinn(2010)AdSpR_45_445}, respectively.
Two nearby background positions (``HB 21N BG'' and ``HB 21S BG,'' Table \ref{tbl-obs}) were also observed in order to inspect if any background \Htwo{} line emission exists.

\begin{figure}
\center{
\includegraphics[scale=0.45,viewport=0 0 500 400]{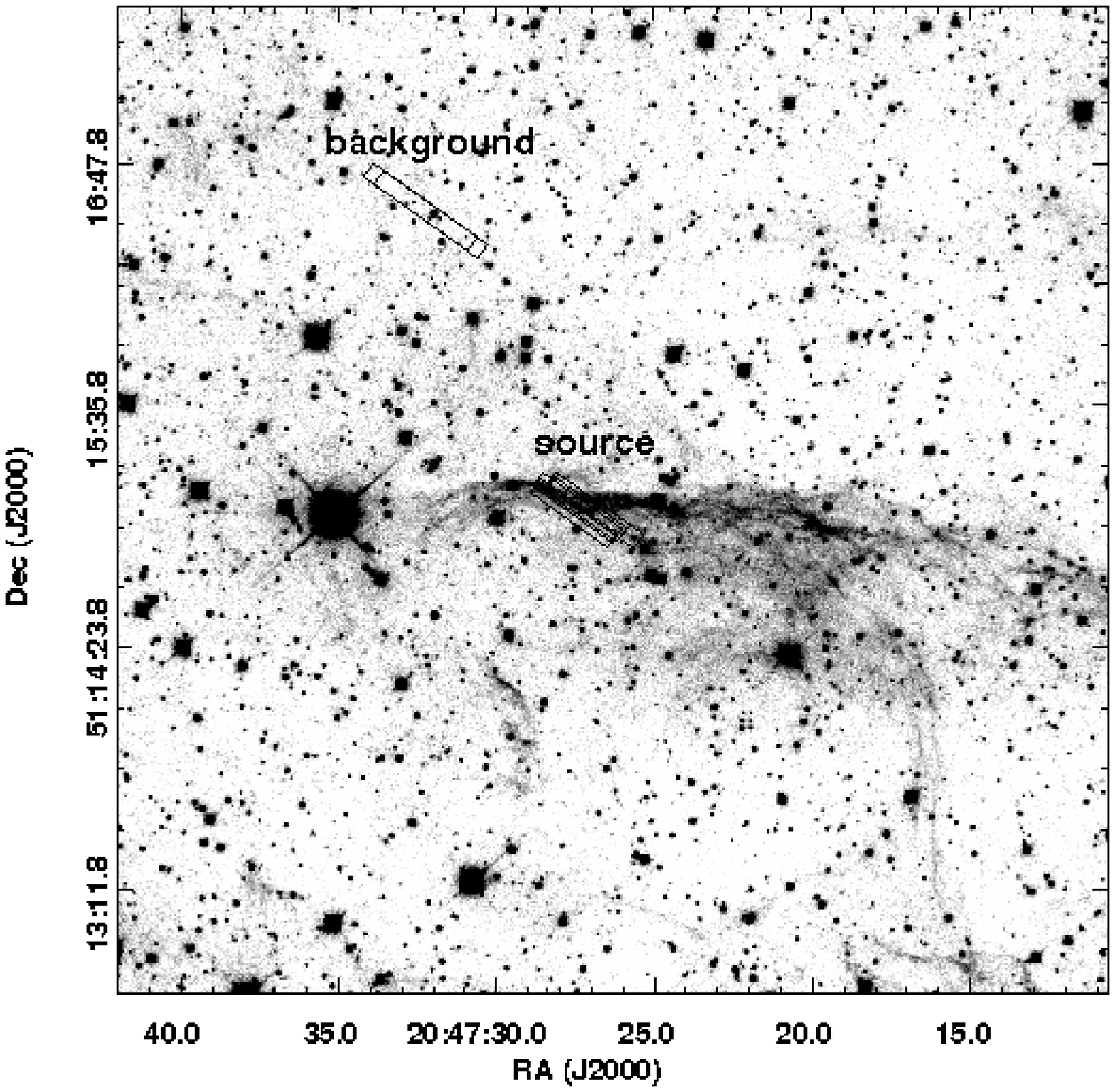}
\includegraphics[scale=0.45,viewport=0 0 500 400]{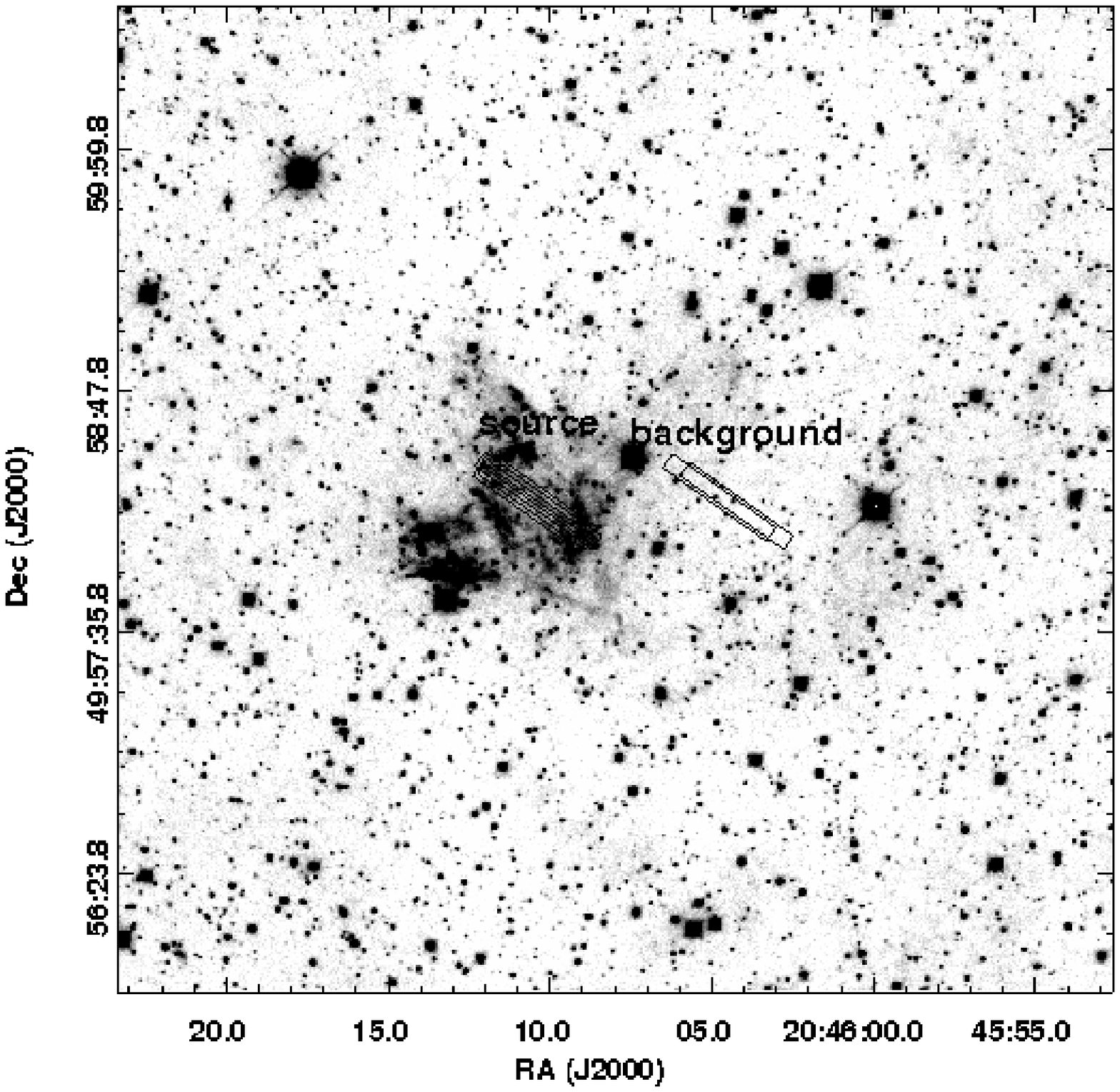}
}
\caption{The observed slit positions for HB 21N (\emph{left}) and HB 21S (\emph{right}). The two regions correspond to ``N2front'' \citep{Shinn(2009)ApJ_693_1883} and ``S1'' \citep{Shinn(2010)AdSpR_45_445}, respectively. The background images are \HtwolineK{} 2.12 \um{} images obtained with WIRCAM of CFHT telescope. The direction to the remnant's center is southward and northward in the \emph{left} and \emph{right} figures, respectively. \label{fig-slit}}
\end{figure}

The observations were carried out with the Infrared Camera \citep[IRC,][]{Onaka(2007)PASJ_59_S401} of the \akari{} satellite \citep{Murakami(2007)PASJ_59_S369}, on 2008 June 1-3 during the post-helium phase.
All the observations were performed with the observation mode IRCZ4, which is designed for general spectroscopic observations.
The IRCZ4 mode acquires the spectra from multiple exposures without dithering, taking one reference image in the middle of the exposures \citep{Onaka(2009)mana}.
Based on this reference image, we corrected the observed astrometric positions employing the 2MASS catalog \citep{Skrutskie(2006)AJ_131_1163}; the resulting astrometric accuracy is smaller than $1.4''$.
We used the $5''\times48''$ slit and the grism, whose spectral resolution and wavelength coverage are $\Delta\lambda\sim0.03$ \um{} and $2.5-5.0$ \um, respectively.
Figure \ref{fig-slit} shows the slit positions over the \HtwolineK{} 2.12 \um{} image of HB 21.
The \Htwo{} 2.12 \um{} image was acquired with the Wide-field InfraRed Camera (WIRCAM) of Canada-France-Hawaii Telescope (CFHT) on 2007 August, and the data reduction was carried out employing the SIMPLE package \citep{Wang(2010)ApJS_187_251}.

\begin{figure}
\center{
\includegraphics[scale=0.8]{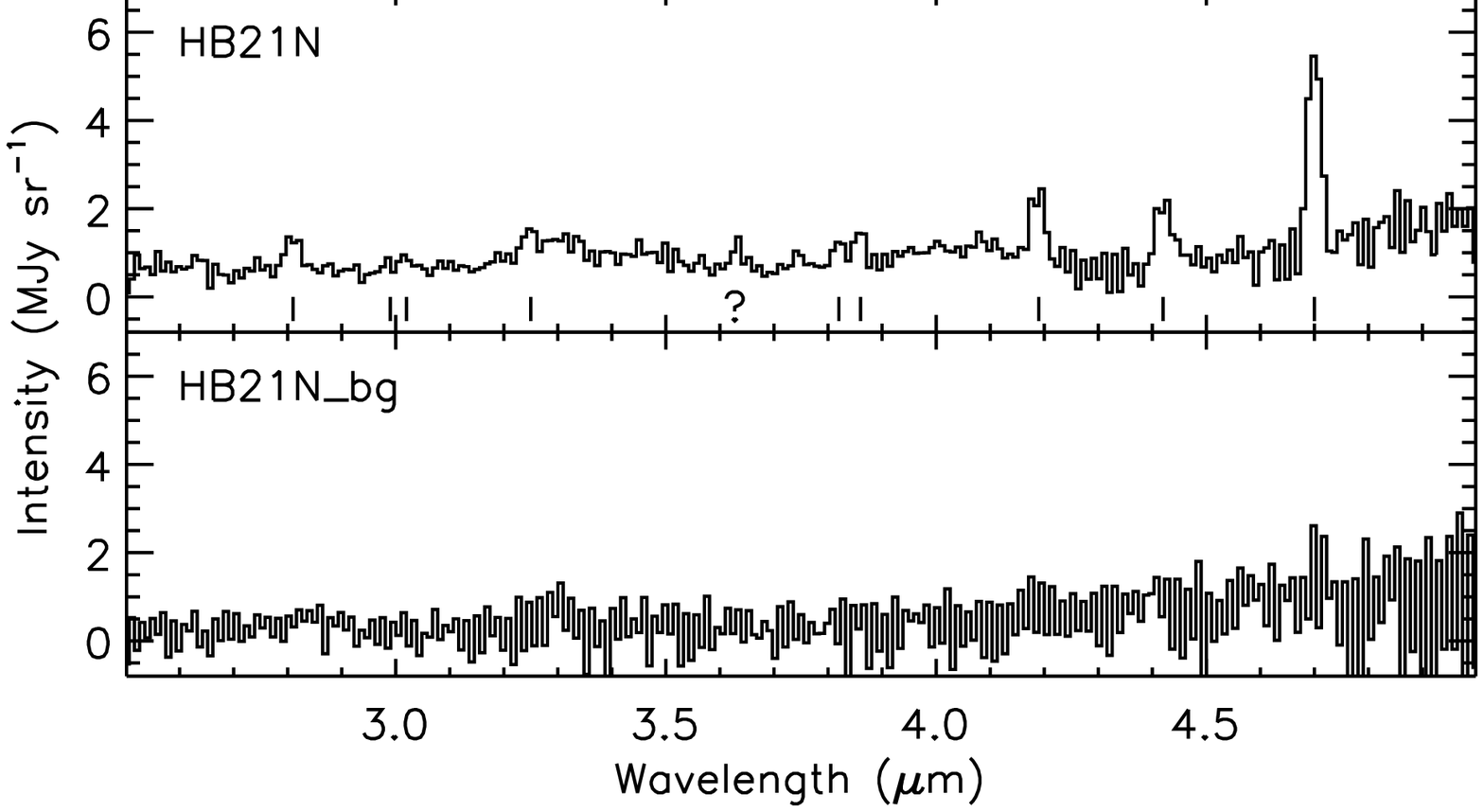}
\includegraphics[scale=0.8]{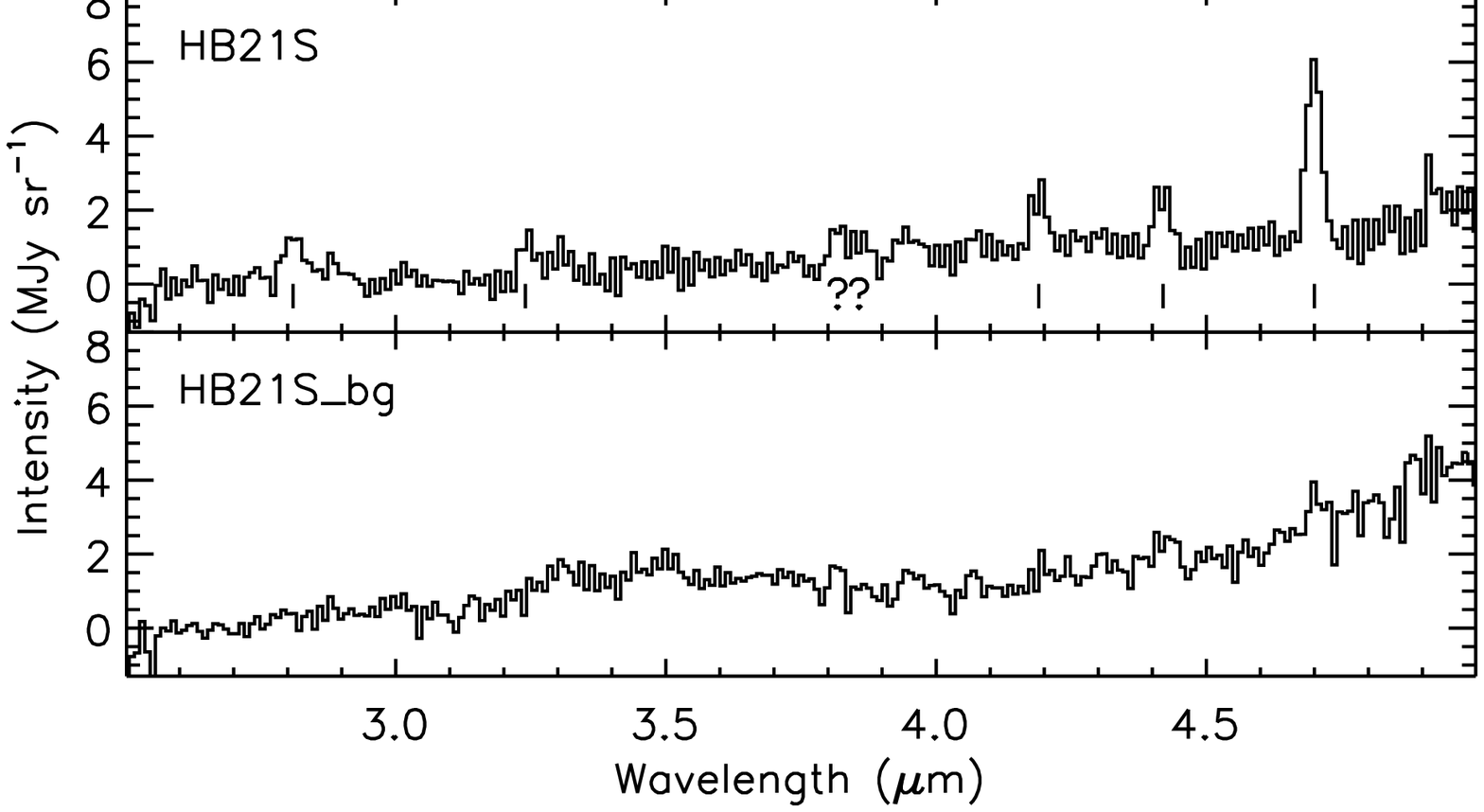}
}
\caption{The \akari{} IRC near-infrared spectra of HB 21N (\emph{top}), HB 21S (\emph{bottom}), and their backgrounds (cf. Fig.~\ref{fig-slit}). The vertical bars indicate the location of \Htwo{} emission lines (cf. Table \ref{tbl-result}). ``?'' indicates the location of doubtful \Htwo{} emission line. For clarity, the error bars are omitted; the errors estimated from the pipeline are shown in Figures \ref{fig-fit-hb21n} and \ref{fig-fit-hb21s}. \label{fig-spec}}
\end{figure}

We first applied the reduction pipeline for the \akari{} post-Helium data \cite[version 20110301,][]{Onaka(2009)mana}.
The column pull-down effect was corrected and the hot pixels were masked out.
No smoothing and tilt-correction were applied to the two dimensional spectral image, while running the pipeline.
Then, we extracted a spectrum for each observation ID (cf. Table \ref{tbl-obs}) by averaging the pixel values along the slit direction.
Again these multiple spectra for each pointing position were averaged together, and the results are presented in Figure \ref{fig-spec}.
Some spectra show an alternating noise pattern along the wavelength bin, but they do not affect the measurement of line intensity we intend.
The calibration systematic error of 10 \% (private communication with the AKARI helpdesk) was squarely summed to the line intensity error (Table \ref{tbl-result}).

\section{Analysis and Results \label{ana-res}}
\subsection{Line Intensity and Level Population \label{ana-res-pop}}
Figure \ref{fig-spec} shows the \akari{} spectra of HB 21.
A few evident emission lines are seen from the spectra of ``HB 21N'' and ``HB 21S.''
On the other hand, no emission line is seen from the spectra of the background positions, ``HB 21N BG'' and ``HB 21S BG''; hence, the background contribution of line emission is negligible.
We identified all the detected emission lines are of \Htwo, comparing the line positions to those detected in IC 443 \citep{Shinn(2011)ApJ_732_124}.

\begin{figure}
\center{
\includegraphics[scale=0.3]{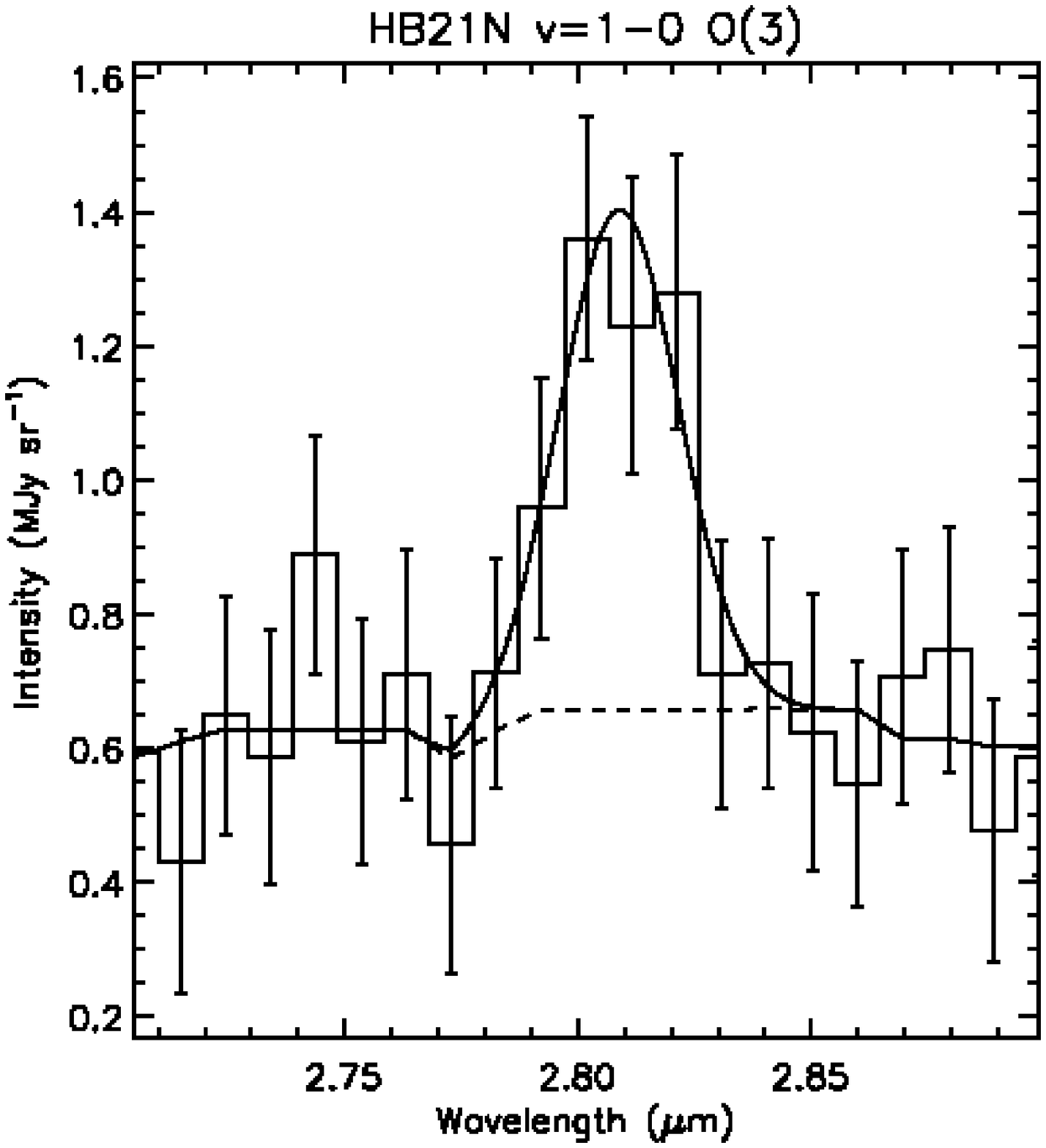}
\includegraphics[scale=0.3]{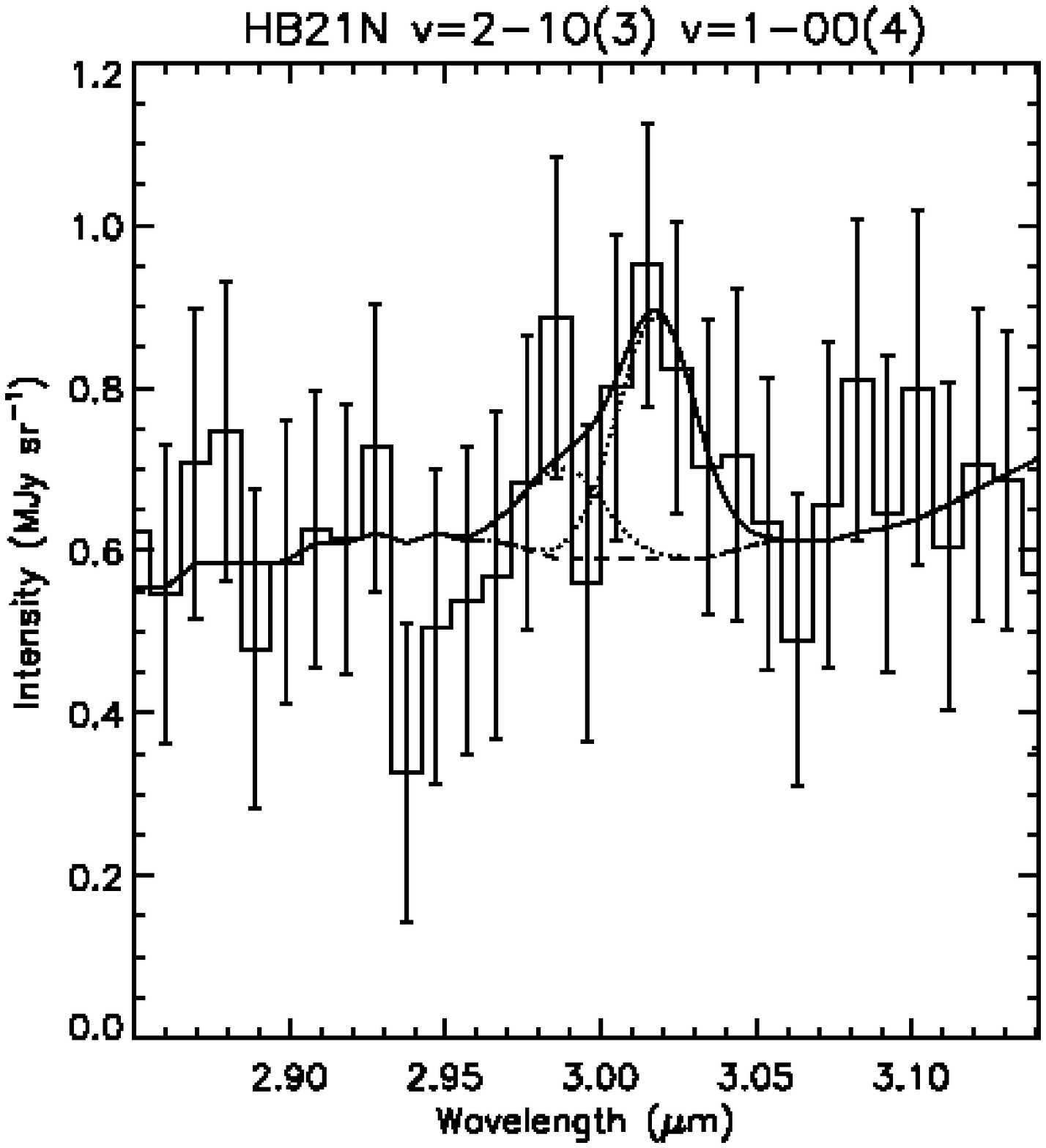}
\includegraphics[scale=0.3]{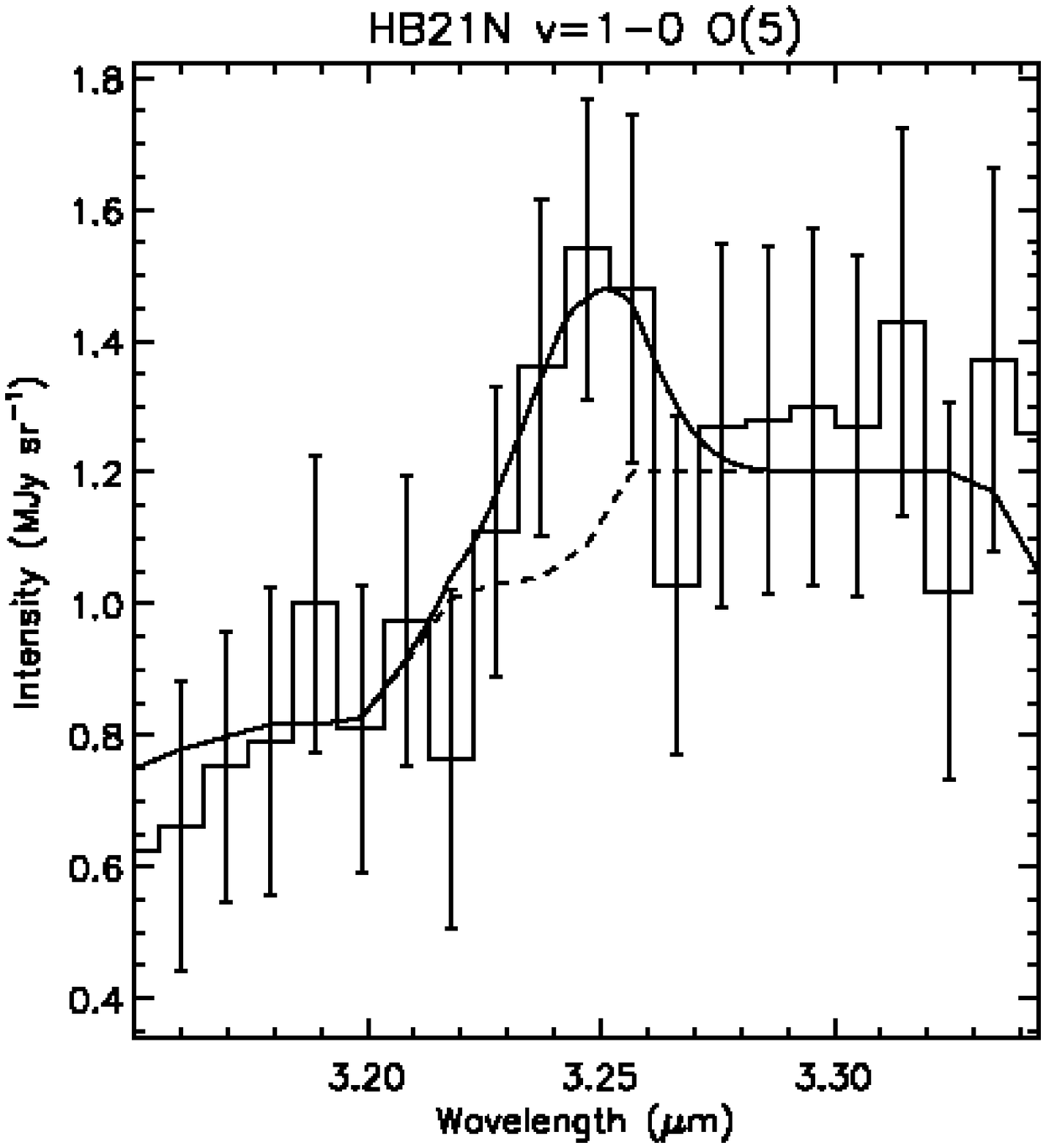}
\includegraphics[scale=0.3]{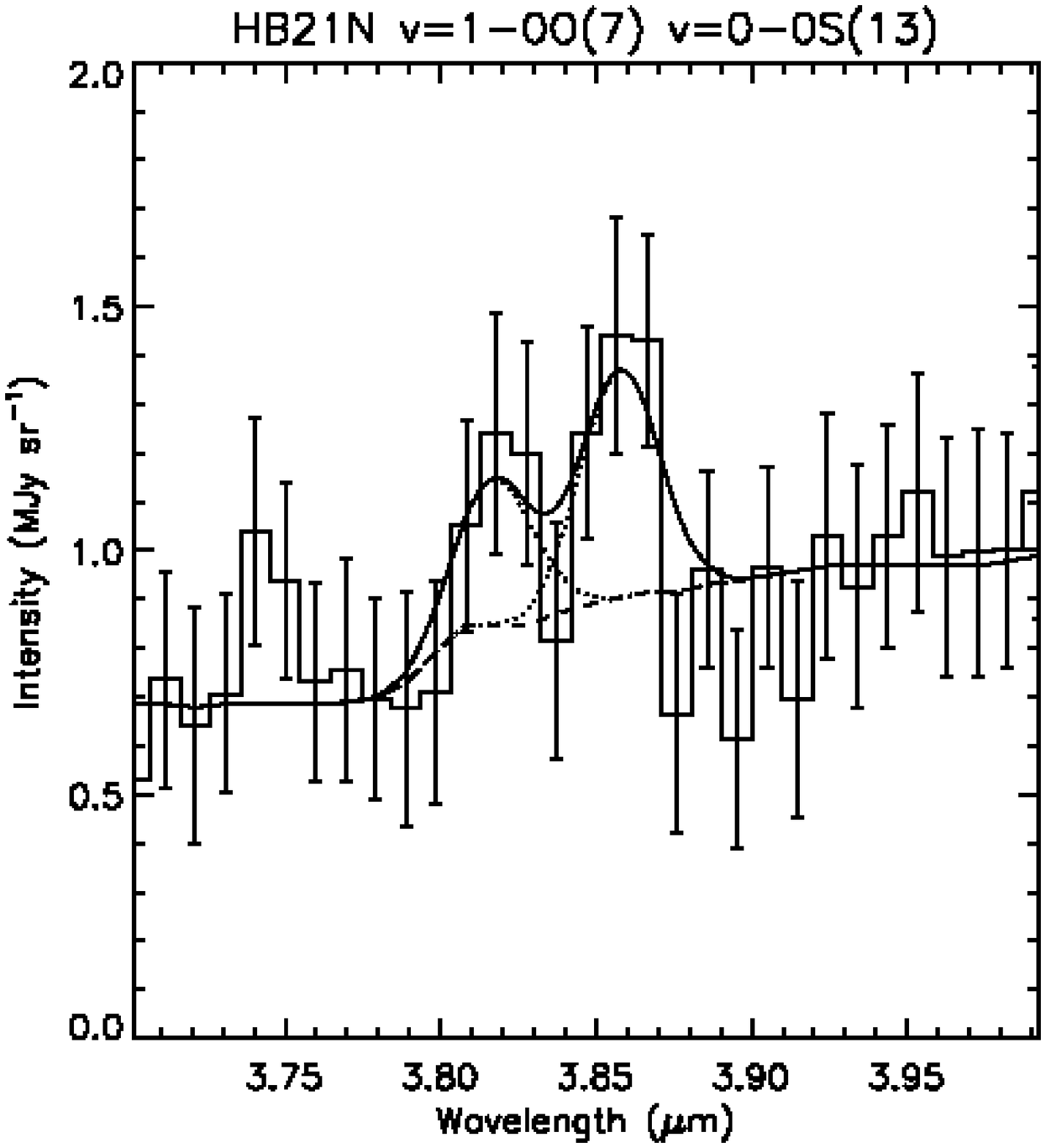}
\includegraphics[scale=0.3]{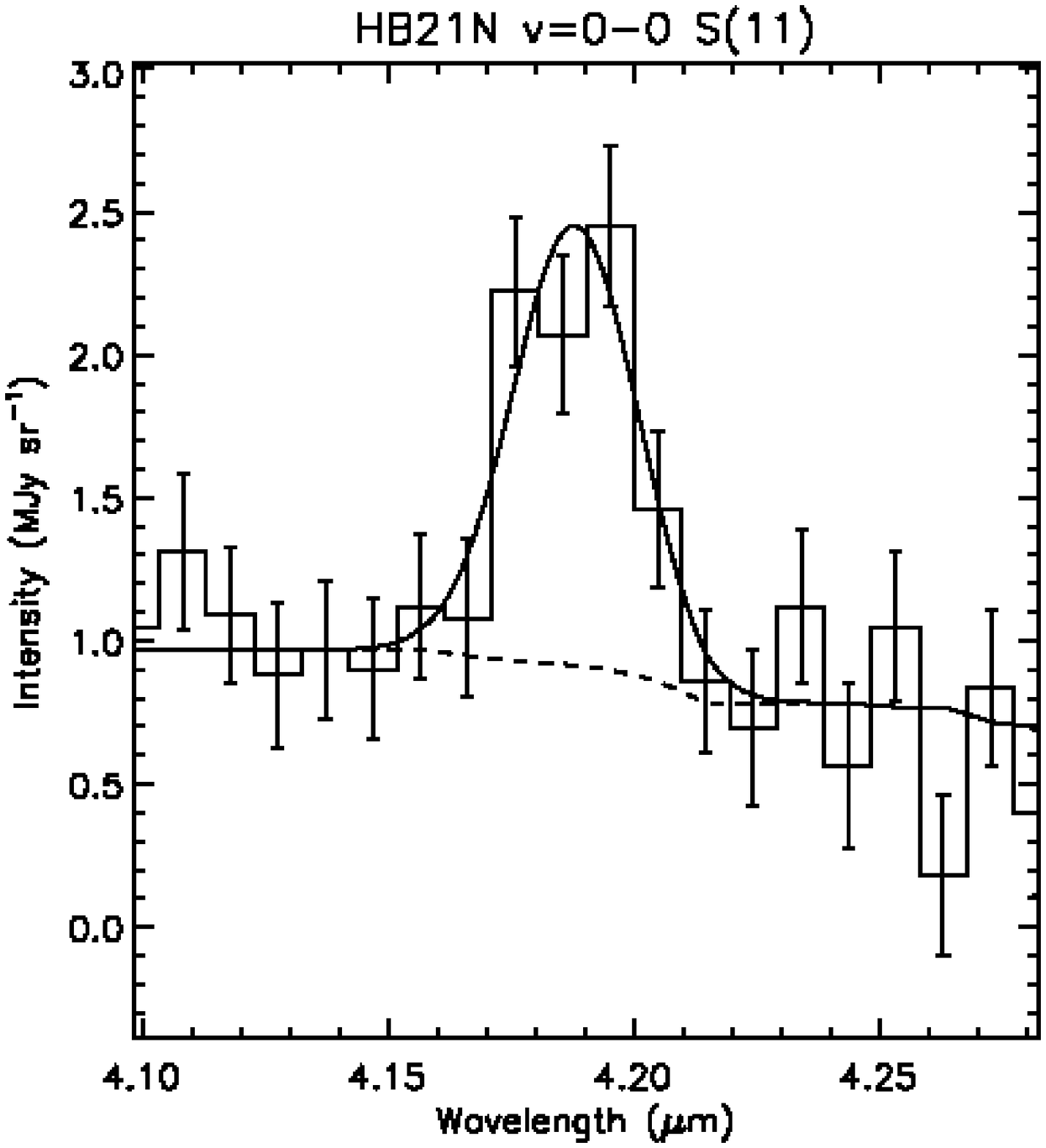}
\includegraphics[scale=0.3]{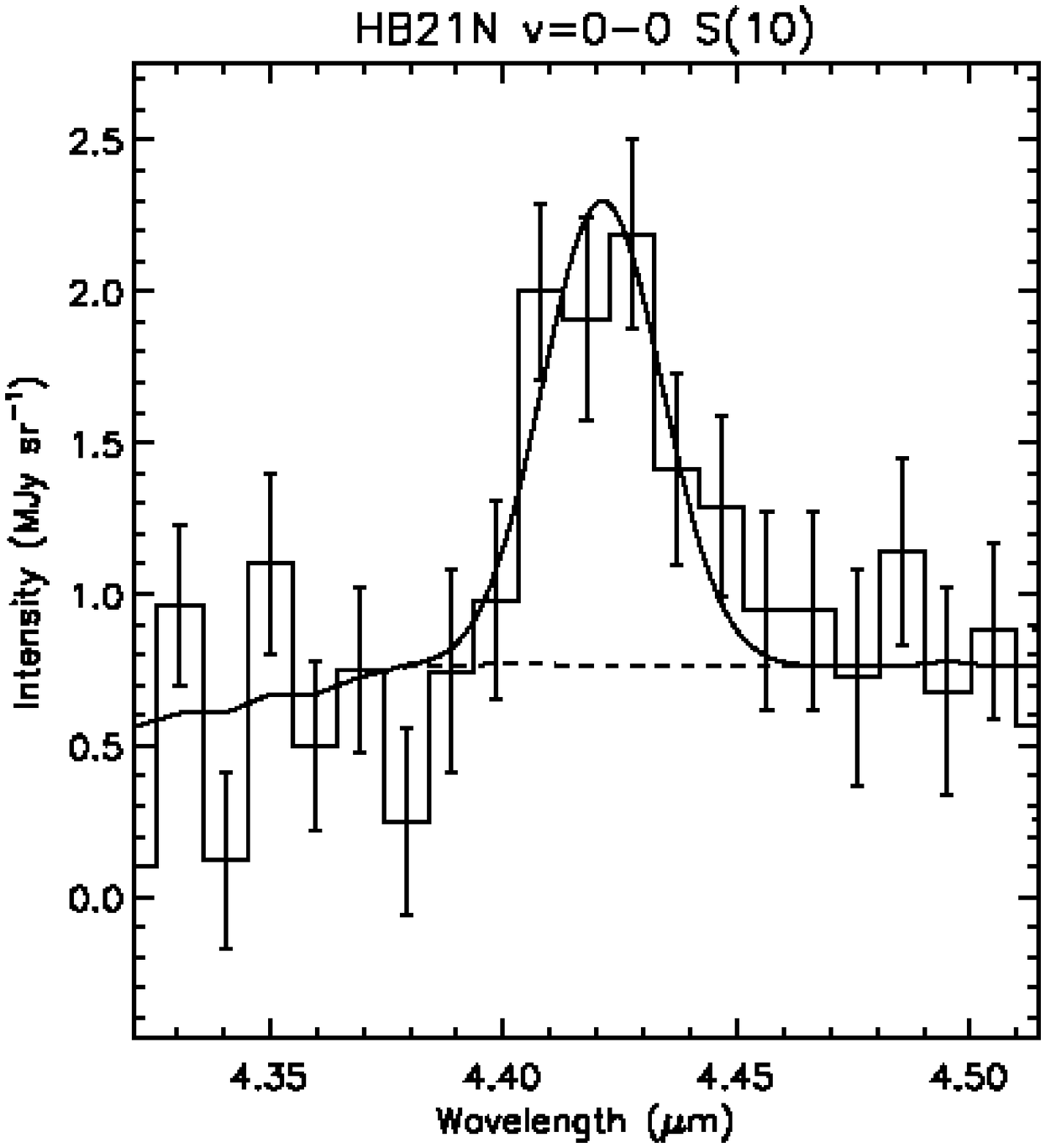}
\includegraphics[scale=0.3]{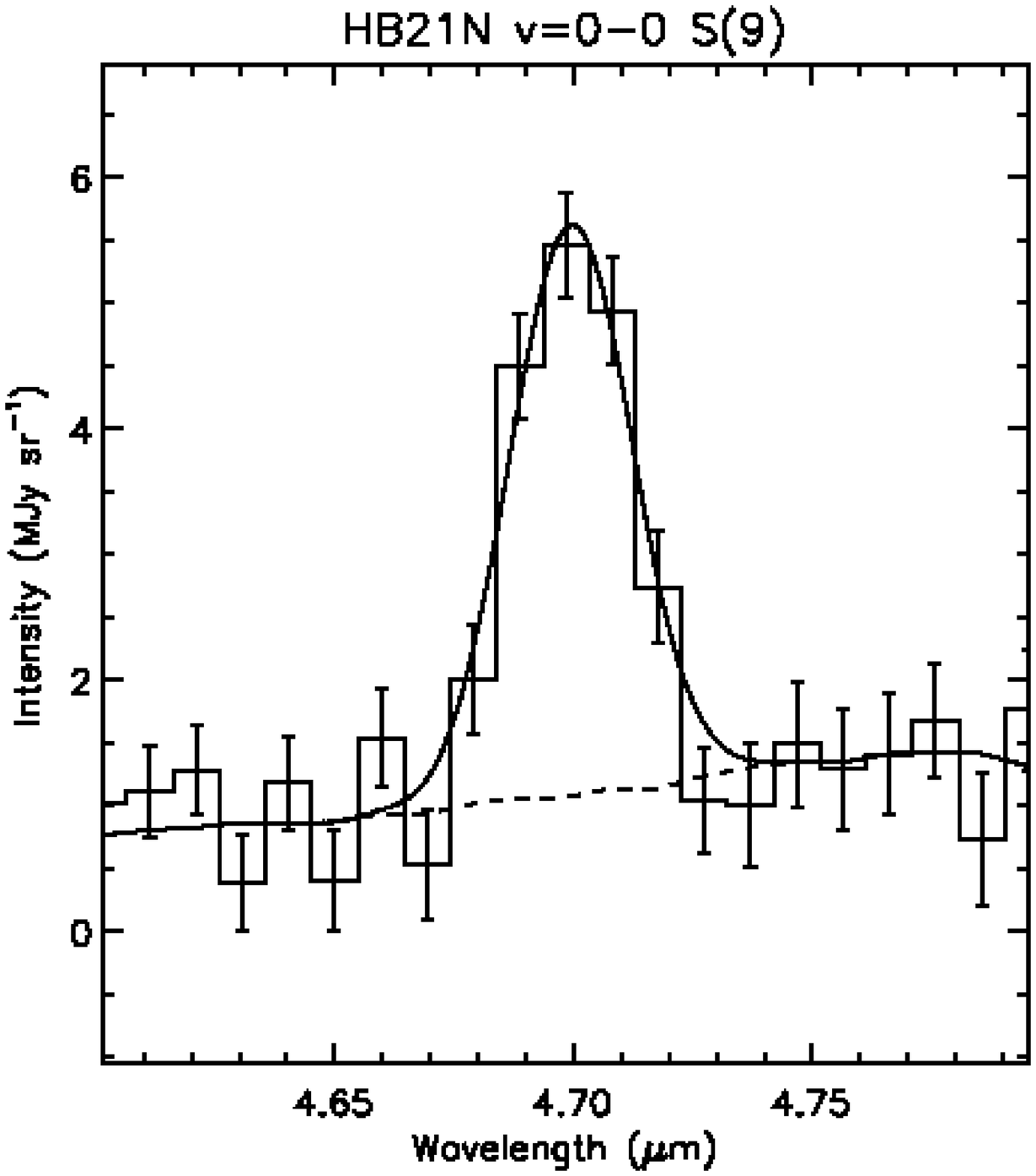}
}
\caption{Fitting for the \Htwo{} emission lines observed toward HB 21N. The \emph{solid} and \emph{dashed} lines indicate the continuum+line and continuum, respectively. The blended line components are indicated by the \emph{dotted} lines, if any. \label{fig-fit-hb21n}}
\end{figure}

\begin{figure}
\center{
\includegraphics[scale=0.3]{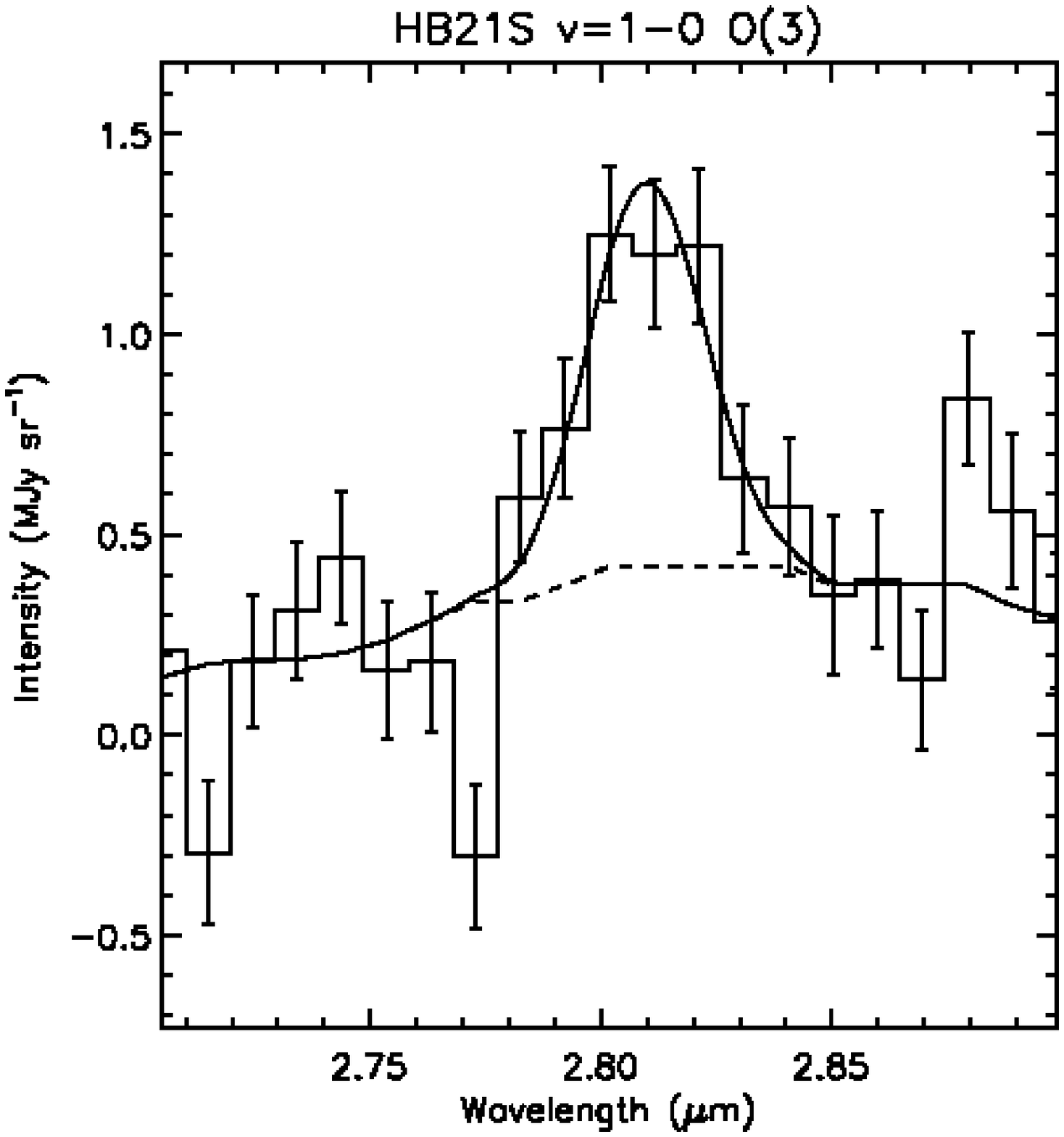}
\includegraphics[scale=0.3]{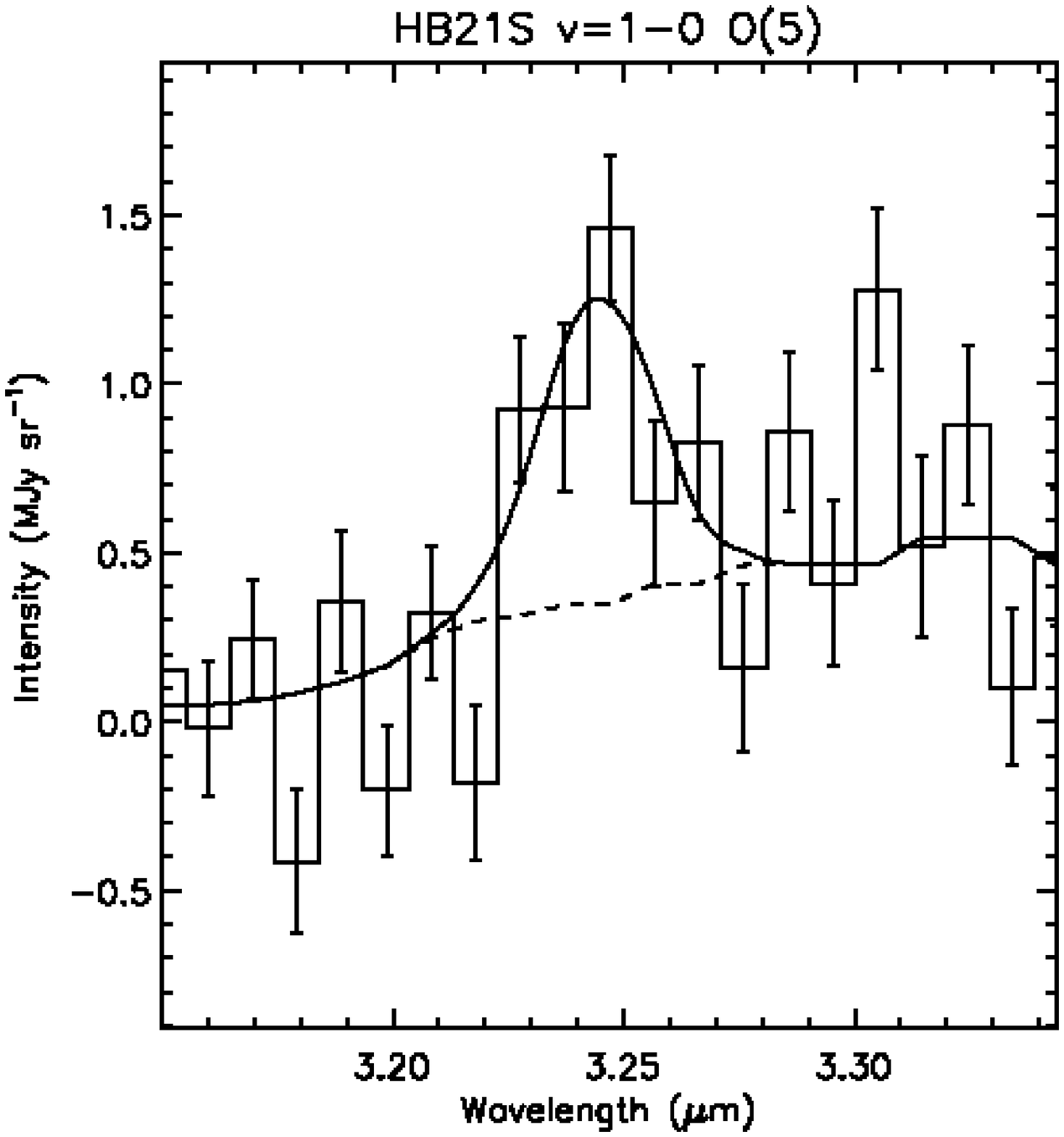}
\includegraphics[scale=0.3]{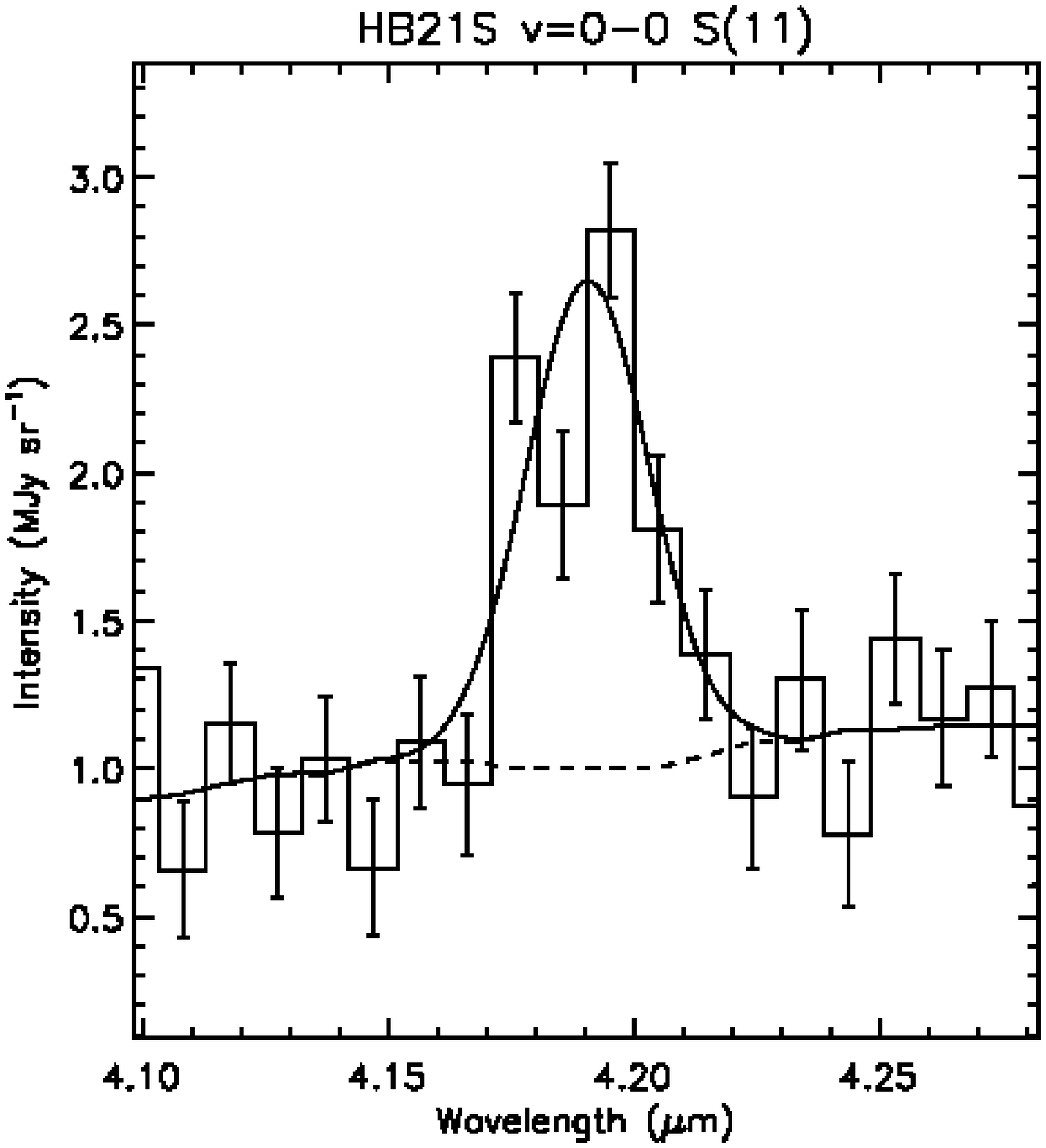}
\includegraphics[scale=0.3]{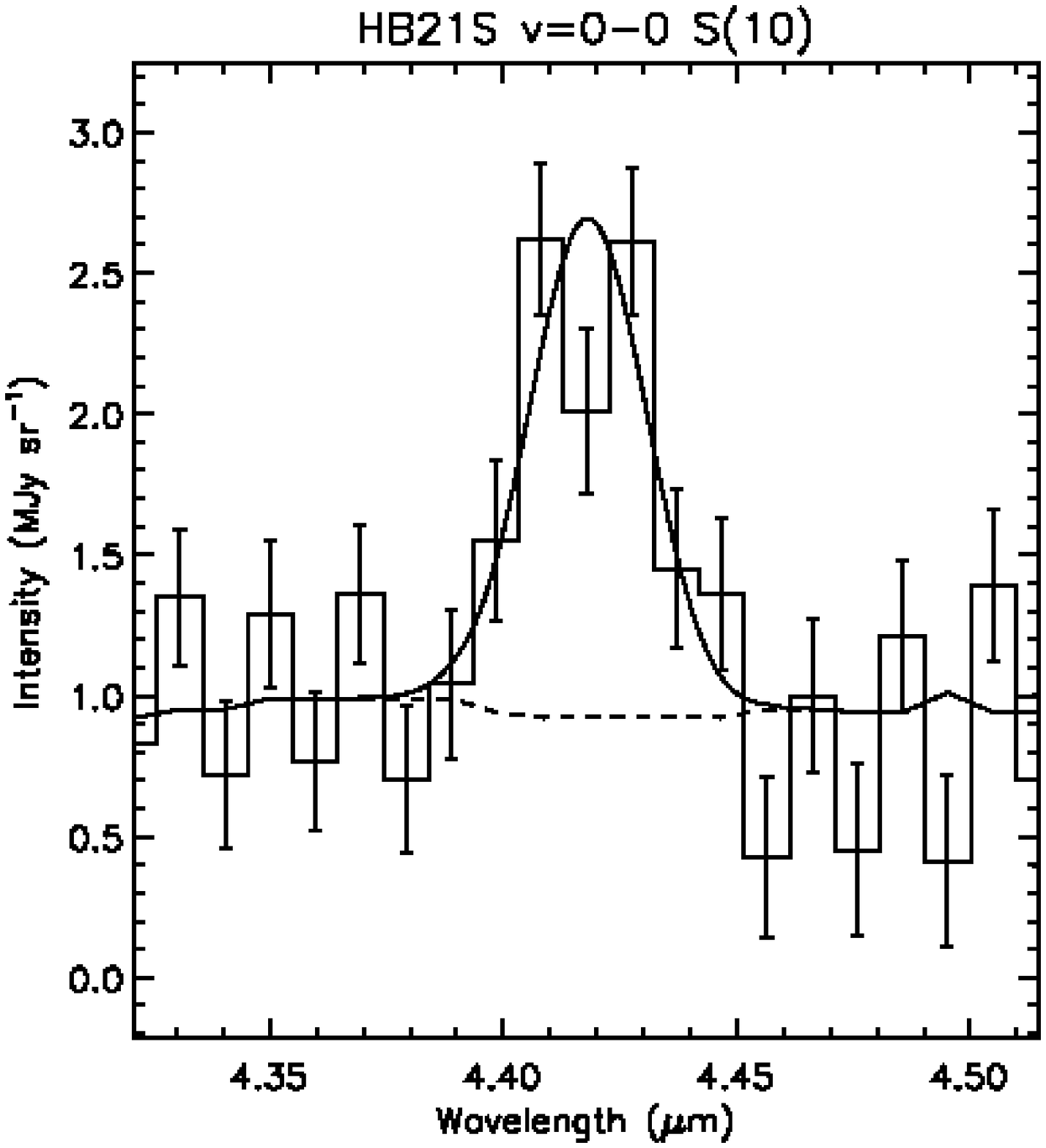}
\includegraphics[scale=0.3]{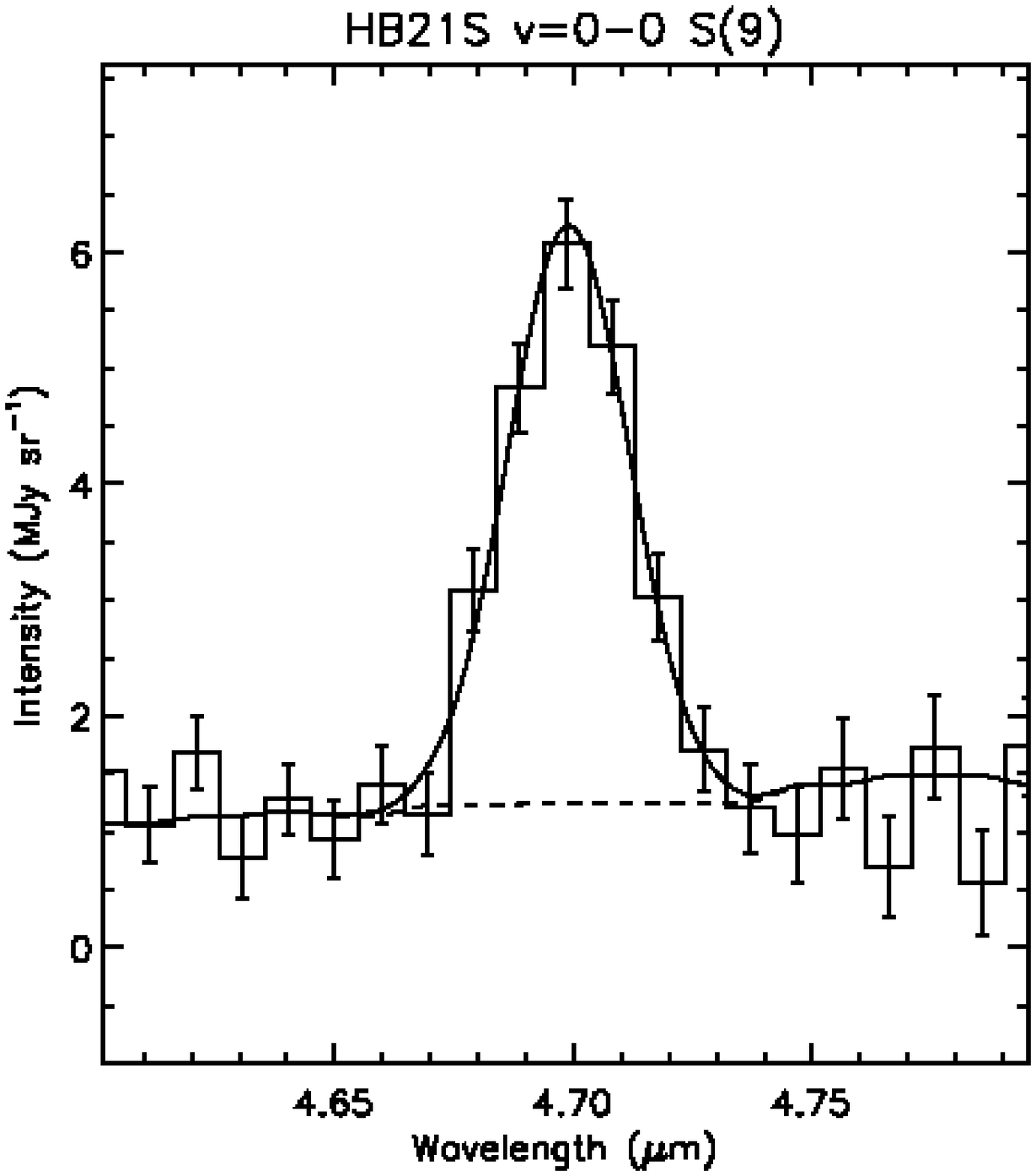}
}
\caption{Fitting for the \Htwo{} emission lines observed toward HB 21S. The rest is the same as Figure \ref{fig-fit-hb21n}. \label{fig-fit-hb21s}}
\end{figure}

The line intensities were measured by fitting a Gaussian profile plus a synthetic continuum to the emission line spectrum.
The synthetic continuum was made by median-smoothing the observed spectrum with a box kernel.
The kernel width is wide enough to smooth the line features out; it is about 0.19 \um.
This continuum is displayed as a dashed line in Figures \ref{fig-fit-hb21n} and \ref{fig-fit-hb21s}.
We set the full-width-at-half-maximum of the Gaussian profile as the spectral resolution of IRC, $\Delta\lambda=0.03$ \um.
When two lines are overlapped, we fitted both lines simultaneously.
The fitting results are shown in Figures \ref{fig-fit-hb21n} and \ref{fig-fit-hb21s}, and the measured line intensities are listed in Table \ref{tbl-result}.

In order to derive the level population of the shocked \Htwo{} gas, we first performed the reddening correction to the measured line intensities.
We employed the extinction curve of ``Milky Way, \Rv$=3.1$'' \citep{Weingartner(2001)ApJ_548_296,Draine(2003)ARA&A_41_241}, and adopted \defNH$=3.5\times10^{21}$ \Ncm{} derived from the X-ray absorption towards the central region of HB 21 \citep{Lee(2001)inproc}.
Then, these reddening-corrected line intensities were used to derive the level population, assuming that the \Htwo{} emission is optically thin.
This assumption is appropriate since the \Htwo{} emission lines become optically thick when \NHtwo{} $\ga10^{24}$ \Ncm, as noted in \cite{Shinn(2011)ApJ_732_124}.
If there was such a high \NHtwo{} $\ga10^{24}$ \Ncm{} toward HB 21, the \akari{} mid-infrared images \citep{Shinn(2009)ApJ_693_1883} should have shown absorbed features against background radiation, because the \NHtwo{} exceeds that of infrared dark clouds \citep{Bergin(2007)ARA&A_45_339}.

\begin{figure}
\center{
\includegraphics[scale=0.45]{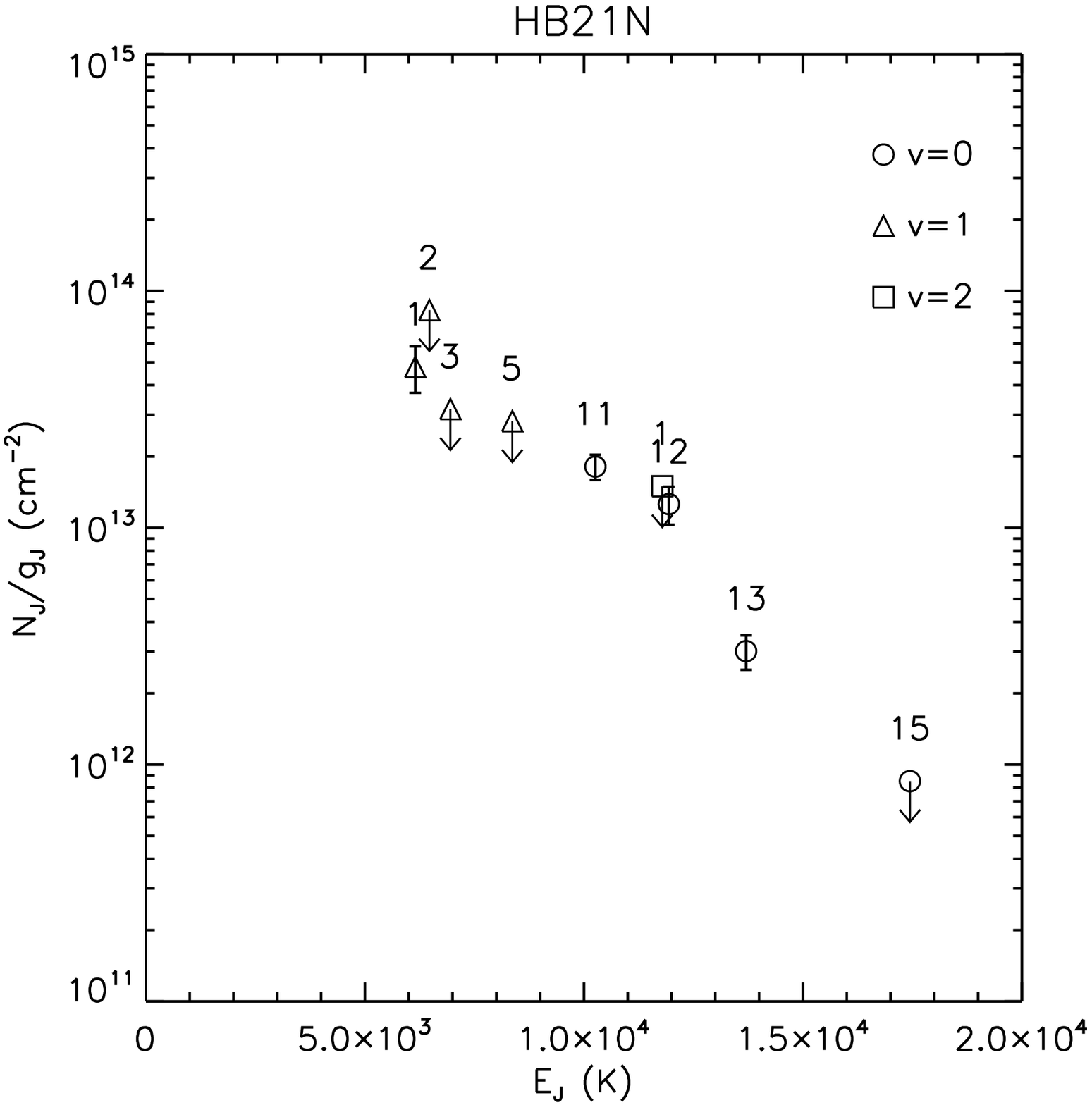}
\includegraphics[scale=0.45]{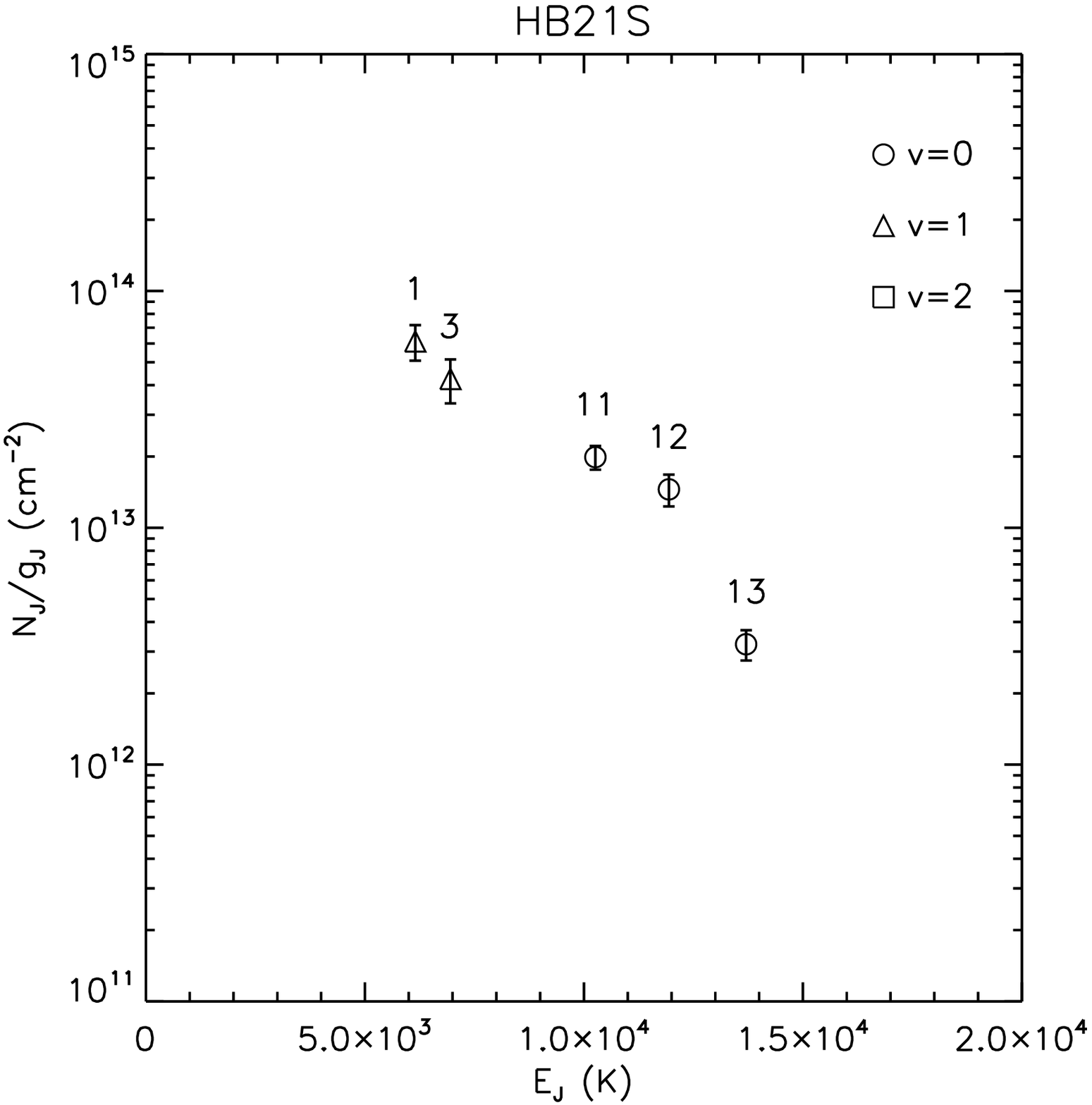}
}
\caption{The extinction-corrected population diagrams for HB 21N (\emph{left}) and HB 21S (\emph{right}). The extinctions were corrected, using \defNH$=3.5\times10^{21}$ \Ncm{} \citep{Lee(2001)inproc} and the extinction curve of ``Milky Way, $R_V=3.1$'' \citep{Weingartner(2001)ApJ_548_296,Draine(2003)ARA&A_41_241}. The rotational quantum number ($J$) is printed out near the corresponding point. \label{fig-pop}}
\end{figure}

We used the following equation to derive the extinction-corrected level population of the shocked \Htwo{} gas.
\begin{equation}
N_{rc}(\upsilon,J)=\frac{4\pi\lambda}{hc}\frac{I_{rc}(\upsilon,J\rightarrow\upsilon',J')}{A(\upsilon,J\rightarrow\upsilon',J')},
\end{equation}
$I_{rc}(\upsilon,J\rightarrow\upsilon',J')$ and $A(\upsilon,J\rightarrow\upsilon',J')$ are the reddening corrected line intensity and the Einstein-A radiative transition probability, respectively.
$\lambda$, $h$, and $c$ are the transition wavelength, Planck constant, and light speed, respectively.
We employed the molecular physical constants tabulated in the simulation code CLOUDY (version C08.00; \citealt{Ferland(1998)PASP_110_761}).
The derived column densities are listed in Table \ref{tbl-h2col}.
Their population diagrams are shown in Figure \ref{fig-pop}; note that the LTE population appears as a straight line in this diagram.

The populations of (\vJ)=(0,11), (0,12), (0,13) show a zigzag pattern.
The OPRs estimated with the single $T_{ex}$ assumption are $1.8\pm0.4$ and $1.7\pm0.3$ for HB 21N and HB 21S, respectively ($T_{ex}\sim2000$ K).
These OPRs are non-equilibrium OPRs (cf.~section \ref{intro}).
Note that IC 443B also showed the zigzag pattern \citep{Shinn(2011)ApJ_732_124}; the OPRs are estimated to be $2.4\pm0.4$ and $2.1\pm0.3$ for (\vJ)=(0,11-13) and (1,1-3), respectively ($T_{ex}\sim1000-2000$ K).
We here underline that these (\vJ) levels all reside at high energy, \EvJ$\ga7000$ K.

In order to ascertain that the zigzag pattern seen from the levels (\vJ)=(0,11-13) reflects the intrinsic OPR $<3.0$, we examined as follows.
We checked the flux calibration procedure including the data reduction pipeline, and found no indication of systematic features that can mimic the zigzag pattern.
We also checked if any blended line exists for $\upsilon=0-0$ S(10) 4.41 \um{}, which corresponds to the level (\vJ)=(0,12).
There is no known strong atomic or ionic lines from shocks around 4.41 \um{} \citep[cf.][]{Allen(2008)ApJS_178_20}.
The line blending between \Htwo{} emission lines is dealt with in section \ref{ana-res-plmod-spec}.

% The line $\upsilon=1-1$ S(11) 4.42 \um{} can be blended with the line $\upsilon=0-0$ S(10).
% The intensity of $\upsilon=1-1$ S(11) is about 10 \% of $\upsilon=0-0$ S(10) intensity in LTE of $T=2000$ K.
% This type of contribution is negligible for IC 443B.
% First, there is no $\upsilon=1-0$ O(9) 4.58 \um{} line, which is stronger than $\upsilon=1-1$ S(11) at LTE T=2000K \citep{Shinn(2011)ApJ_732_124}.
% Second, the OPR of IC 443B seen from the levels (\vJ)=(1,1-3) is less than 3.0.
% This means that the OPR at \EvJ$\ga7000$ K is likely less than 3.0, which makes the strength of $\upsilon=1-1$ S(11) relative to $\upsilon=0-0$ S(10) even smaller.
% The contribution of $\upsilon=1-1$ S(11) would also be negligible for HB 21N and HB 21S, as the absence of $\upsilon=1-0$ O(9) shows (Fig.~\ref{fig-spec}).
% The $\upsilon=1-1$ S(11) line may make some contributions for the clumps C and G of IC 443 (i.e.,~IC 443C and IC 443G), as the $\upsilon=1-0$ O(9) line was observed \citep{Shinn(2011)ApJ_732_124}.
% However, its amount would be limited up to about 10 \% as mentioned above.

\subsection{Power-law Thermal Admixture Model of Shocked \Htwo{} Gas with a Non-equilibrium Ortho-to-Para Ratio \label{ana-res-plmod}}
The power-law thermal admixture model has been used to describe the level population of shocked \Htwo{} gas \citep[][and references therein]{Shinn(2011)ApJ_732_124}.
We employed the same model used in \cite{Shinn(2011)ApJ_732_124}, and slightly modified it for the reproduction of the non-equilibrium OPR, i.e.,~the zigzag pattern in the population diagram.
The relative population between ortho- and para-levels is controlled by scaling the rate coefficients for the reactive collisions with H, which mediate the ortho-to-para and para-to-ortho transitions \citep[cf.][]{LeBourlot(1999)MNRAS_305_802}.
The final level population is obtained by integrating the following equation,
\begin{eqnarray} \label{eq:pow-mod}
dN &=& a T^{-b} dT \\
a &=& \frac{\mathrm{N}(H_2; T>1000 \mathrm{K})(b-1)}{T_{min}^{1-b}-T_{max}^{1-b}}
\end{eqnarray}
where $a$ and $b$ are constants and $T_{min}=1000$ K and $T_{max}=4000$ K.
The statistical equilibrium was assumed and the collisional partners are \Htwo, He, and H.
We assumed \nHe$=0.2\,$\nHtwo, and the abundance ratio of H to \Htwo{} was set as a free parameter $X_H$.

\begin{figure}
\center{
\includegraphics[scale=0.45]{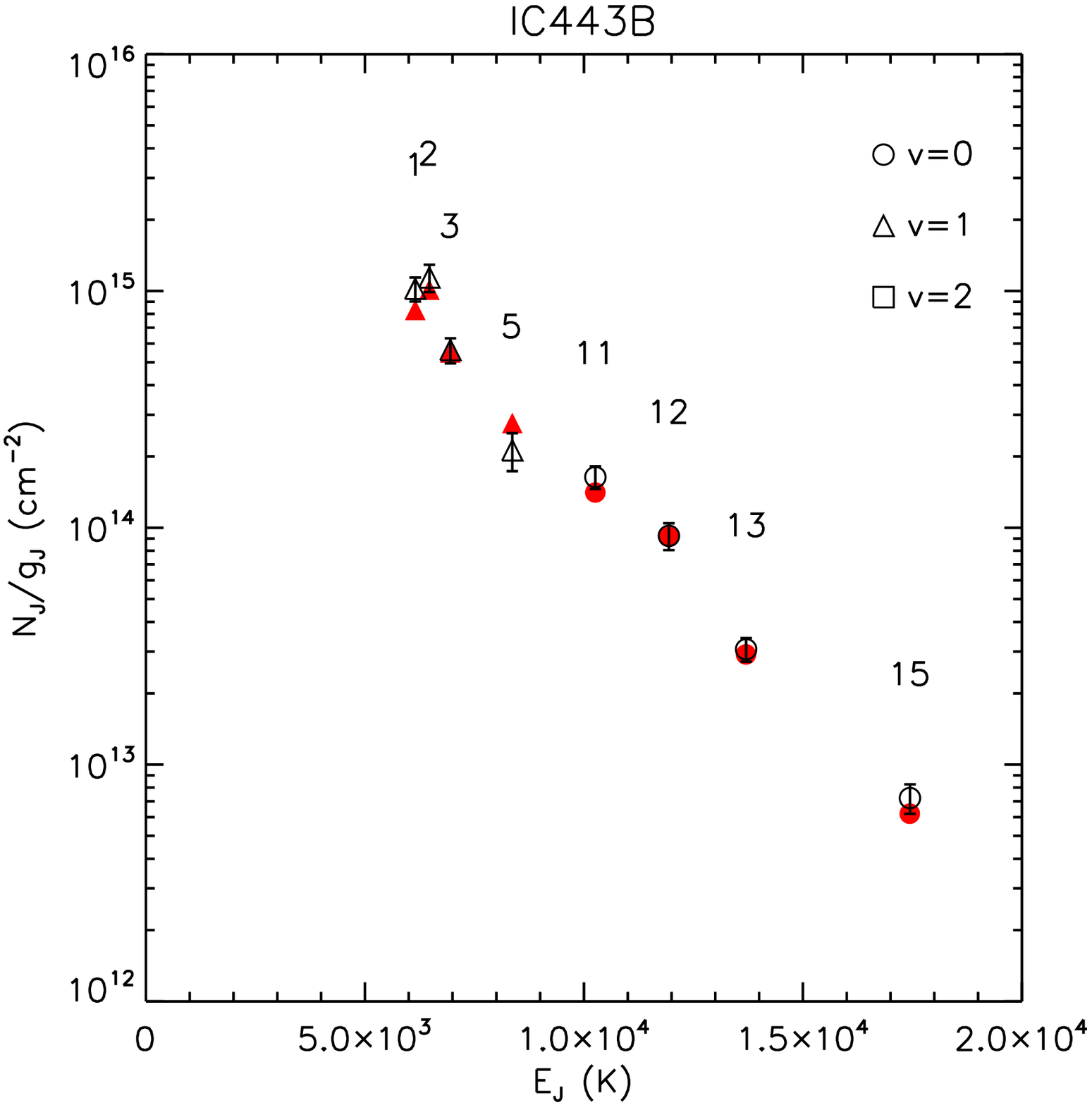}
\includegraphics[scale=0.45]{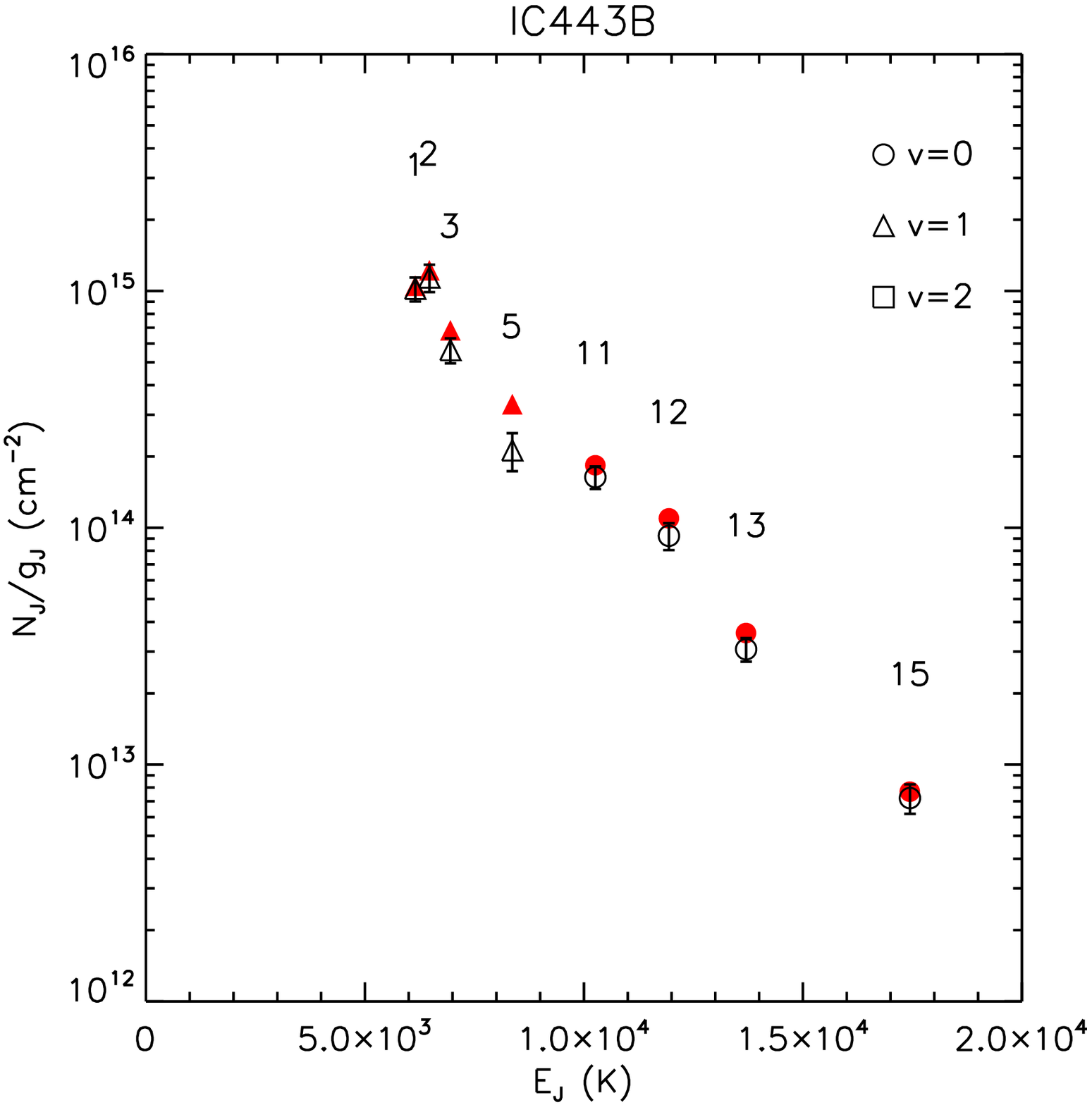} \\
\includegraphics[scale=0.45]{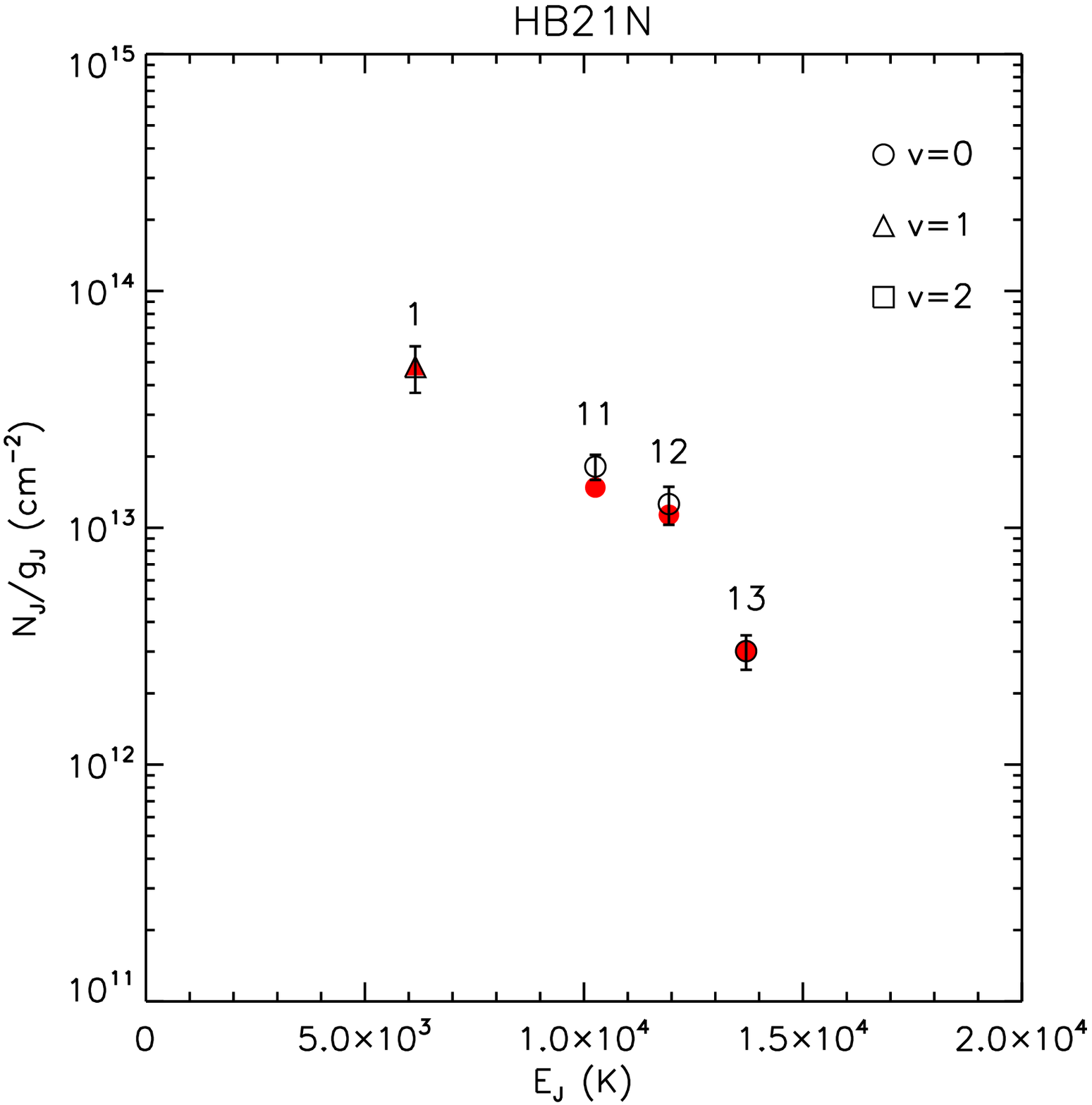}
\includegraphics[scale=0.45]{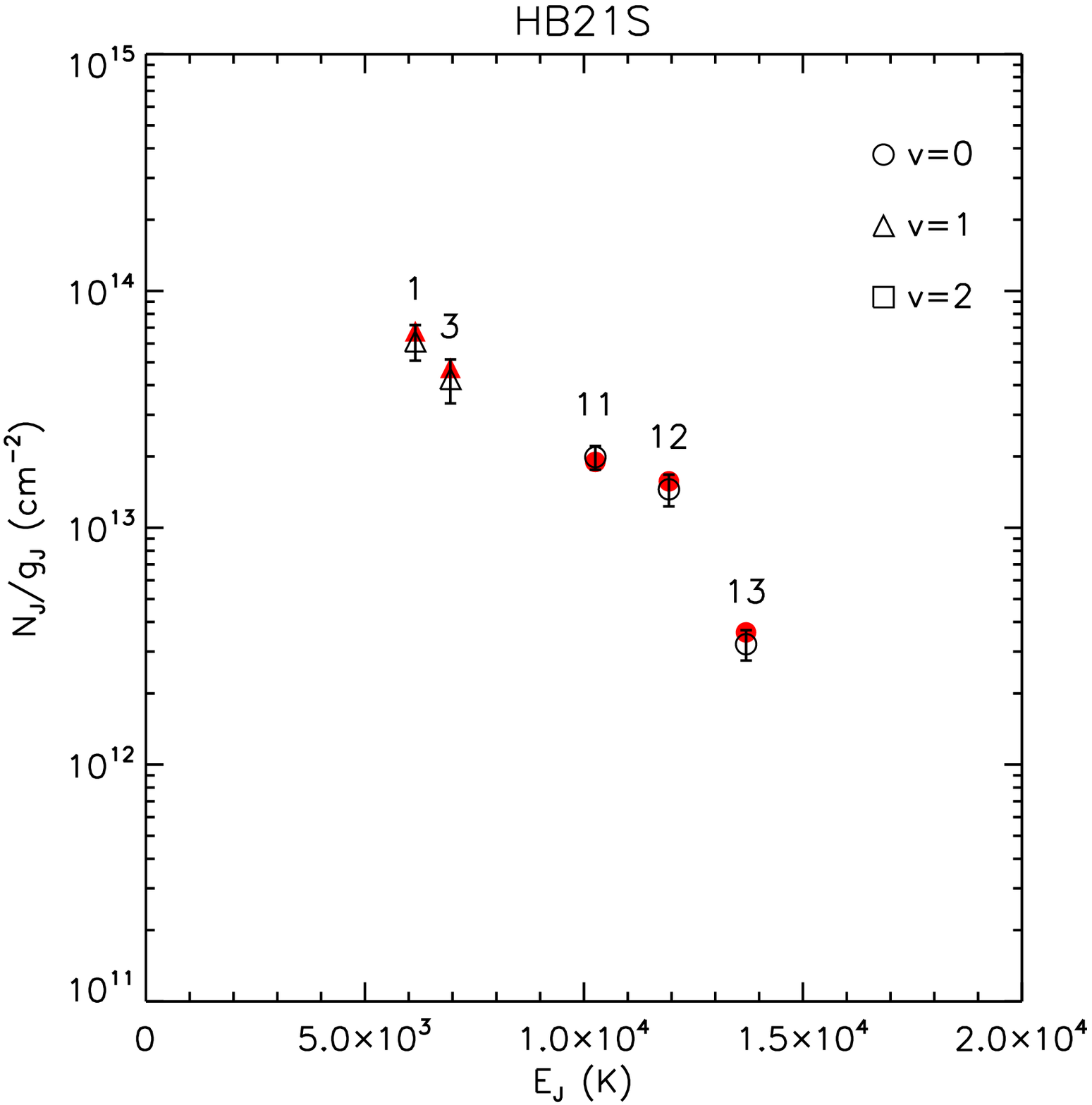}
}
\caption{The model fitting results for the \Htwo{} populations observed in IC 443B (\emph{top}), HB 21N (\emph{bottom-left}), and HB 21S (\emph{bottom-right}). The IC 443B cases are shown for two different parameter settings: X$_H=-1.7$ (\emph{top-left}) and X$_H=-2.5$ (\emph{top-right}) (cf. Table \ref{tbl-mfit}). The filled-red and open-black symbols indicate the model and data points, respectively. (See text for the model description.) The rotational quantum number ($J$) is printed out near the corresponding point. \label{fig-mfit}}
\end{figure}

Based on the previous observations of \cite{Shinn(2011)ApJ_732_124}, we adopted $T_{min}=1000$ K instead of 100 K unlike previous applications \citep[e.g.,][]{Shinn(2009)ApJ_693_1883,Shinn(2010)AdSpR_45_445}.
\cite{Shinn(2011)ApJ_732_124} found that two different densities are required to describe the level population of shocked \Htwo{} gas in the range of $0\la$ \EvJ{} $\la25,000$ K: one with \nHtwo{} of $\sim10^{2.8-3.8}$ \ncm{} for \EvJ{} $\la7000$ K and the other of $\sim10^{5.4-5.8}$ \ncm{} for \EvJ{} $\ga7000$ K.
From these results, they recognized that only $T\ge1000$ K gas is crucial to describe the level population of \EvJ{} $\ga7000$ K.
Therefore, we adopted $T_{min}=1000$ K for the model calculation, since our data points spread over \EvJ{} $\ga7000$ K (Figure \ref{fig-mfit}). 

\begin{figure}
\center{
\includegraphics[scale=0.45]{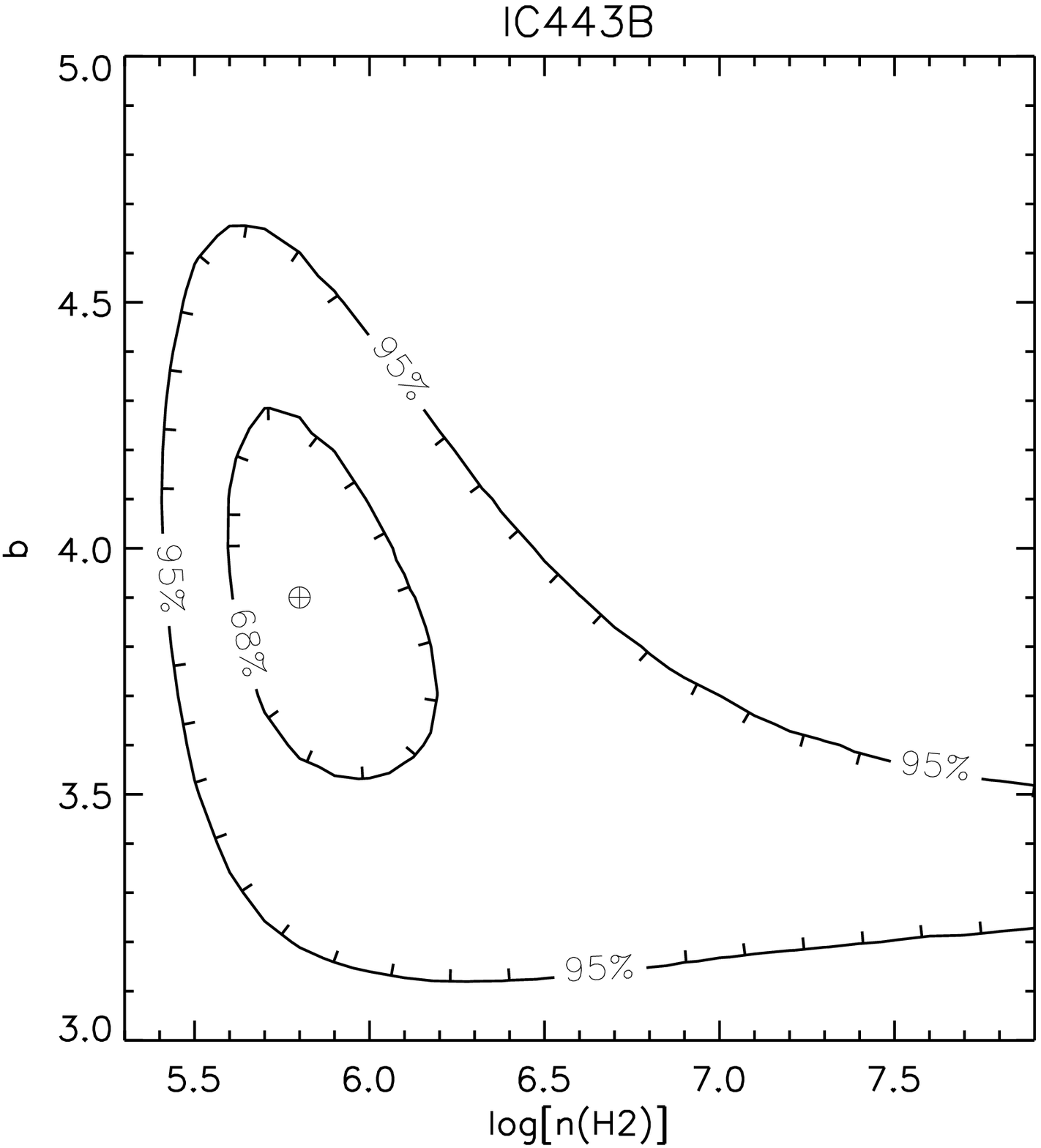}
\includegraphics[scale=0.45]{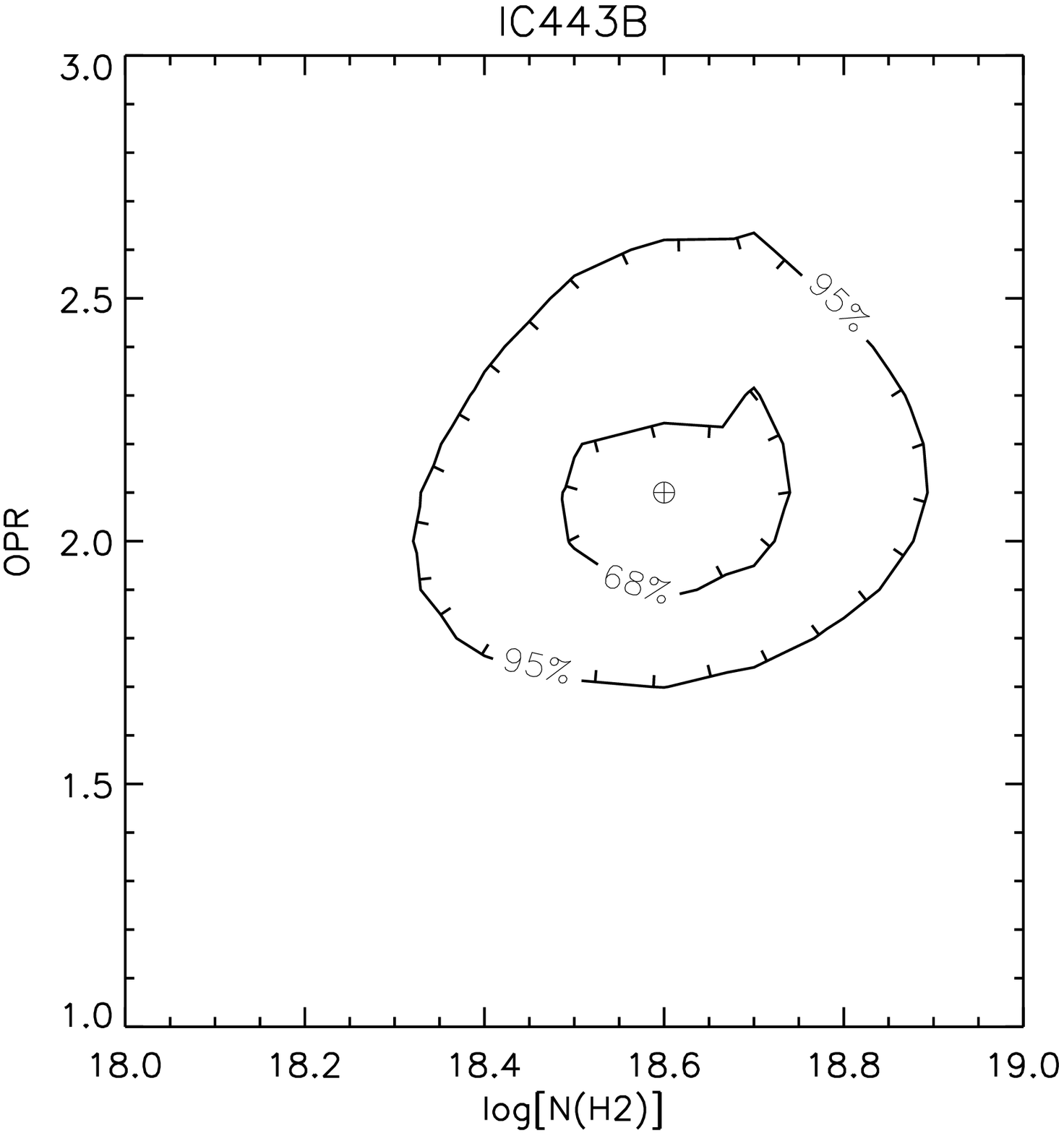}
\includegraphics[scale=0.45]{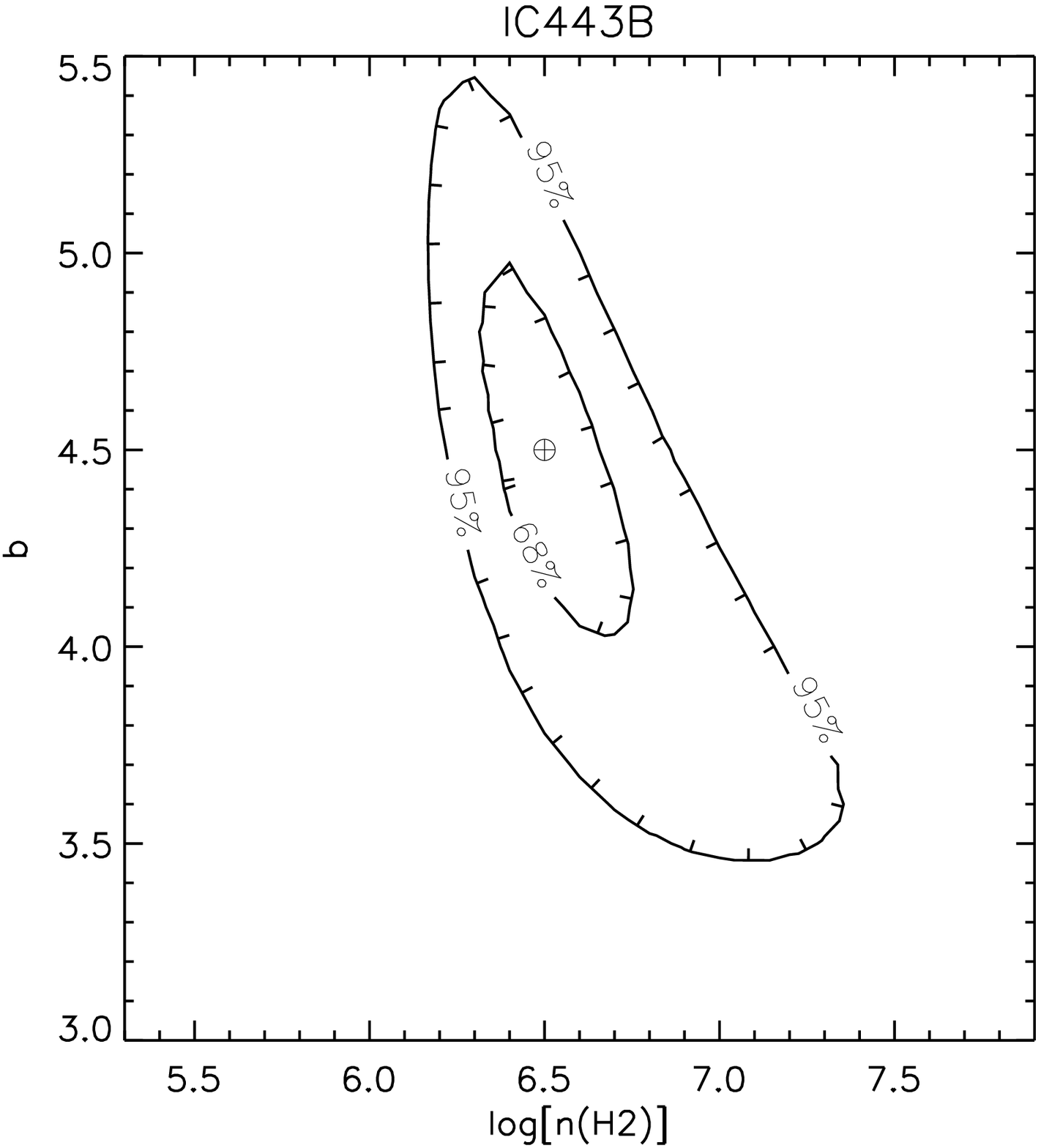}
\includegraphics[scale=0.45]{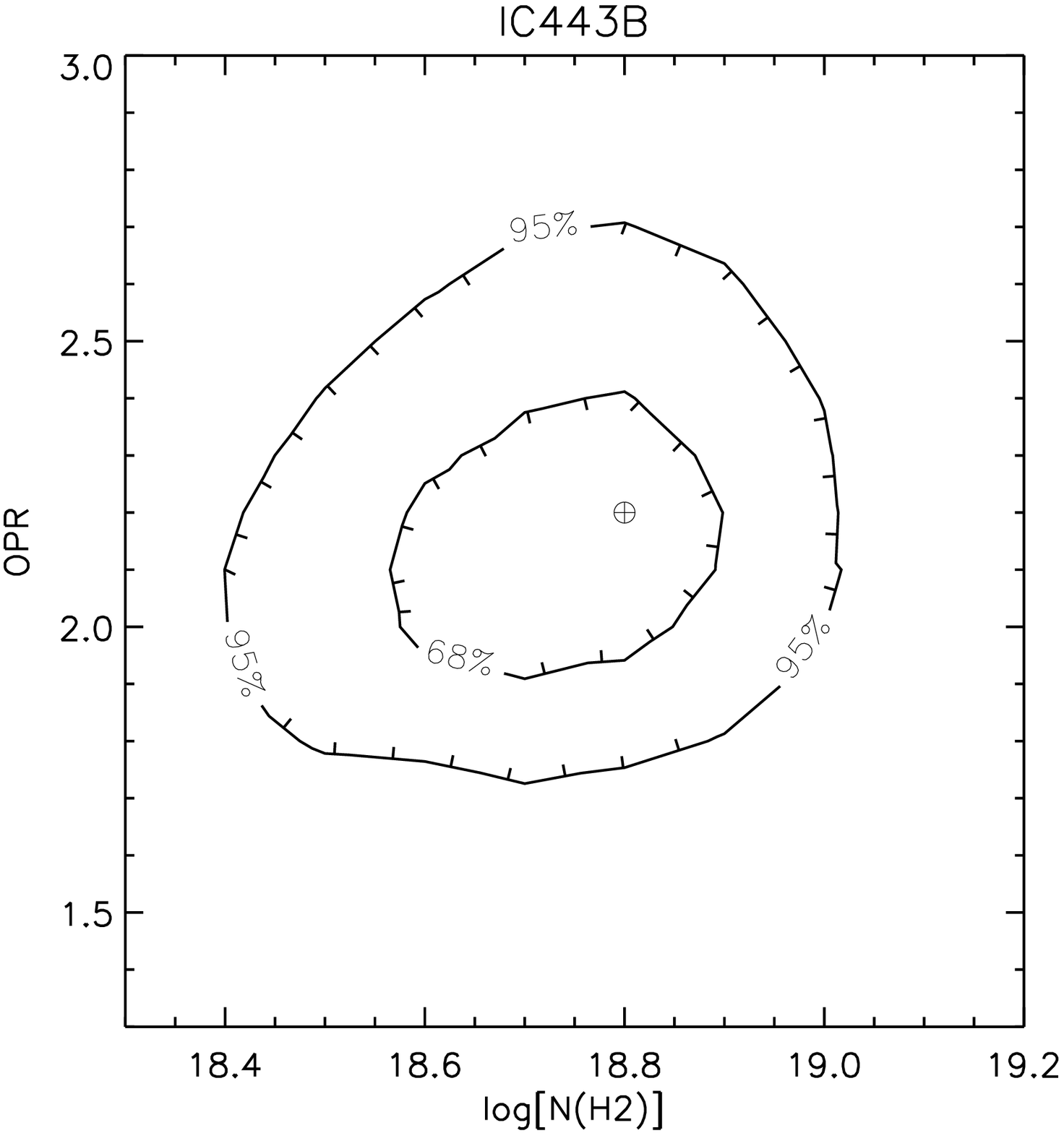}
}
\caption{The contour of $\chi^2$ in the plane of model parameters for IC 443B: X$_H=-1.7$ (\emph{top}) and X$_H=-2.5$ (\emph{bottom}) (cf. Table \ref{tbl-mfit} and Fig.~\ref{fig-mfit}). The 68\% and 95\% confidence levels are outlined. The tick marks along the contours indicate the directions that $\chi^2$ values are decreasing. `$\oplus$' indicates the model parameter values whose $\chi^2$ is minimum; i.e., the best fit. \label{fig-mconf}}
\end{figure}

We then applied the model to the reddening-corrected level population of shocked \Htwo{} gas in IC 443B, HB 21N, and HB 21S.
The fitting results are shown in Figure \ref{fig-mfit}, and the corresponding model parameters are listed in Table \ref{tbl-mfit}.
In the case of IC 443B, we first tried to fit the data without any fixed parameters, but failed to have inversely diverging \nHtwo{} and $X_H$; \nHtwo{} increases over $10^7$ \ncm{} and $X_H$ decreases below $-3.0$.
These two parameters seem to be not well-constrained by the given data and error.
Since such a high \nHtwo{} is unreasonable, we fixed the parameter $X_H$.
Two different fittings were performed with $X_H=-1.7$ and $X_H=-2.5$.
$X_H=-1.7$ is chosen since it was obtained from the analysis of IC 443C \citep{Shinn(2011)ApJ_732_124}.
$X_H=-2.5$ is chosen since IC 443B seems to be experiencing a weaker shock than IC 443C, where $X_H=-1.7$ was obtained.
The intensity ratio $I(\mathrm{CO};J=3-2)/I(\mathrm{CO};J=2-1)$ is lower at IC 443B than IC 443C \citep{Xu(2011)ApJ_727_81}, and the ``W-shaped'' \Htwo{} morphology of IC 443 suggests that IC 443B may be denser than IC 443C \citep[see Fig.~1 in][]{Shinn(2011)ApJ_732_124}.
In the case of HB 21N and HB 21S, we fitted the data fixing both \nHtwo{} and $X_H$, because the data points are too few.
We just show the level population can be described by the power-law thermal admixture model.
We adopted \nHtwo{} $=10^{5.5}$ \ncm{} and $X_H=-2.0$, which are thought to be typical values for shocked \Htwo{} gas of IC 443 \citep[cf.][]{Shinn(2011)ApJ_732_124}.
Figure \ref{fig-mconf} shows the $\chi^2$ contour of the model parameters for IC 443B.
We did not perform the $\chi^2$ contour scan for HB 21N and HB 21S, since the data points are too few, compared to the number of model parameters.

IC 443B shows \nHtwo{} $=10^{5.8-6.5}$ \ncm{} somewhat higher than that of IC 443C, \nHtwo{} $=10^{5.4}$ \ncm{} \citep{Shinn(2011)ApJ_732_124}.
This density difference may be related with the ``W-shaped'' \Htwo{} morphology mentioned above.
\NHtwo{} of IC 443B is $10^{18.6-18.8}$ \Ncm; here, \NHtwo{} means $N$(\Htwo; $T>1000$ K), i.e., the column density of \Htwo{} gas whose temperature is higher than 1000 K.
IC 443B shows $b=3.9-4.5$ larger than that of IC 443C, $b=1.6$.
This difference is natural because $b$ of IC 443C is mainly determined by the data of \EvJ{} $<7000$ K, while it is not for $b$ of IC 443B.
The obtained OPRs are $2.1-2.2$.
Interestingly, the OPR of IC 443B shows a smaller variation than the other parameters under the different adoptions of $X_H$ (cf.~Table \ref{tbl-mfit}).
These OPRs are similar to those estimated from the levels (\vJ)=(0,11-13) and (1,1-3) assuming a single $T_{ex}$: $2.4\pm0.4$ and $2.1\pm0.3$.
The $\chi^2$ contour (Figure \ref{fig-mconf}) shows that there is a weak anti-correlation between \nHtwo{} and $b$, but no correlation between \NHtwo{} and the OPR.
Figure \ref{fig-mconf} definitely shows the non-equilibrium OPR even in 95\% confidence level.
%The upper limit for the (\vJ)=(0,14) level, which is newly measured in this paper, is also compatible with the non-equilibrium OPR as seen from Figure \ref{fig-mfit}.

HB 21N and HB 21S show \NHtwo{} of $10^{17.4}$ \Ncm{} and $10^{17.7}$ \Ncm, respectively.
This is about an order of magnitude smaller than that of IC 443C, which explains the consistent difference in the line intensities \citep[Figure \ref{fig-spec},][]{Shinn(2011)ApJ_732_124}.
HB 21N and HB 21S show $b$ values of 2.6 and 3.3, smaller than IC 443B; this difference is hardly meaningful because the number of HB 21 data points is just half of IC 443B.
The OPRs are 1.8 and 1.6 for HB 21N and HB 21S, respectively.
These values are similar to those estimated from the levels (\vJ)=(0,11-13) assuming a single $T_{ex}$: $1.8\pm0.4$ and $1.7\pm0.3$.

\subsubsection{Spectral Fitting to the Observed Spectrum \label{ana-res-plmod-spec}}
\begin{figure}
\center{
\includegraphics[scale=0.8]{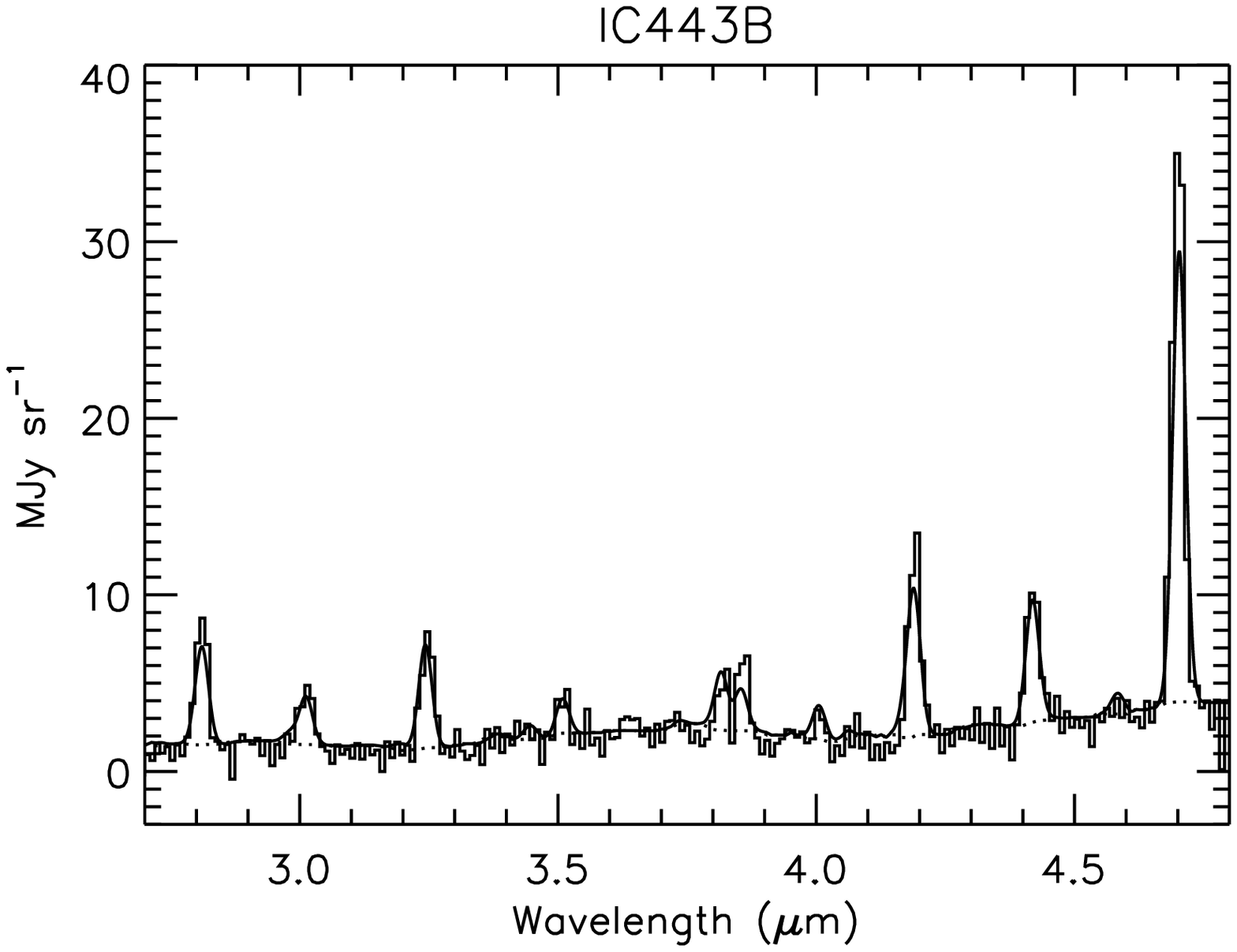}
\includegraphics[scale=0.8]{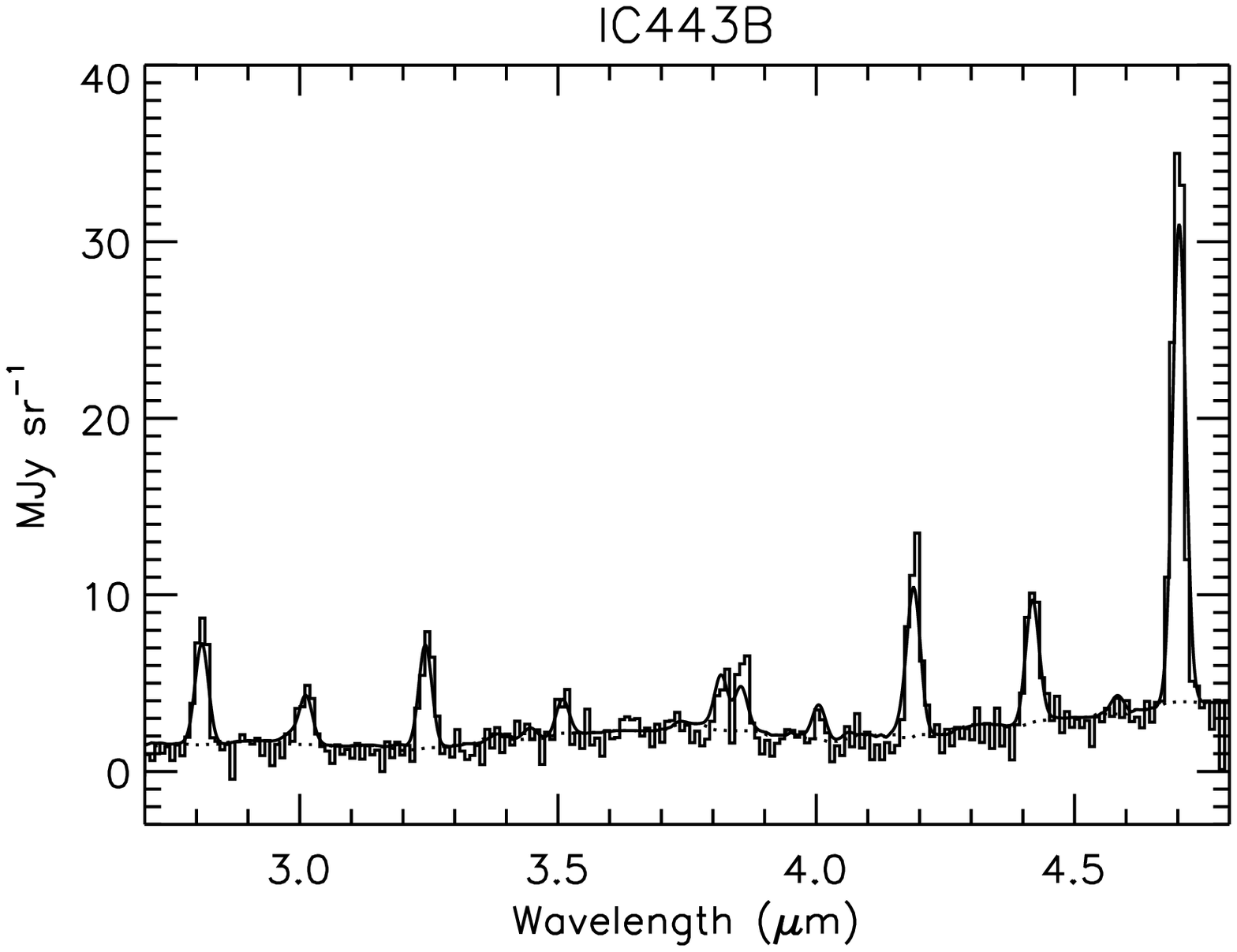}
}
\caption{The model fitting results for the \akari-IRC spectrum observed in IC 443B. The fittings are done for two different parameter settings: X$_H=-1.7$ (\emph{top}) and X$_H=-2.5$ (\emph{bottom}) (cf. Table \ref{tbl-mfit-spec}). The \emph{histogram}, \emph{solid-line}, and \emph{dotted-line} indicate the spectrum data, the fitting result, and the adopted continuum, respectively. \label{fig-mfit-spec}}
\end{figure}

\begin{figure}
\center{
\includegraphics[scale=0.45]{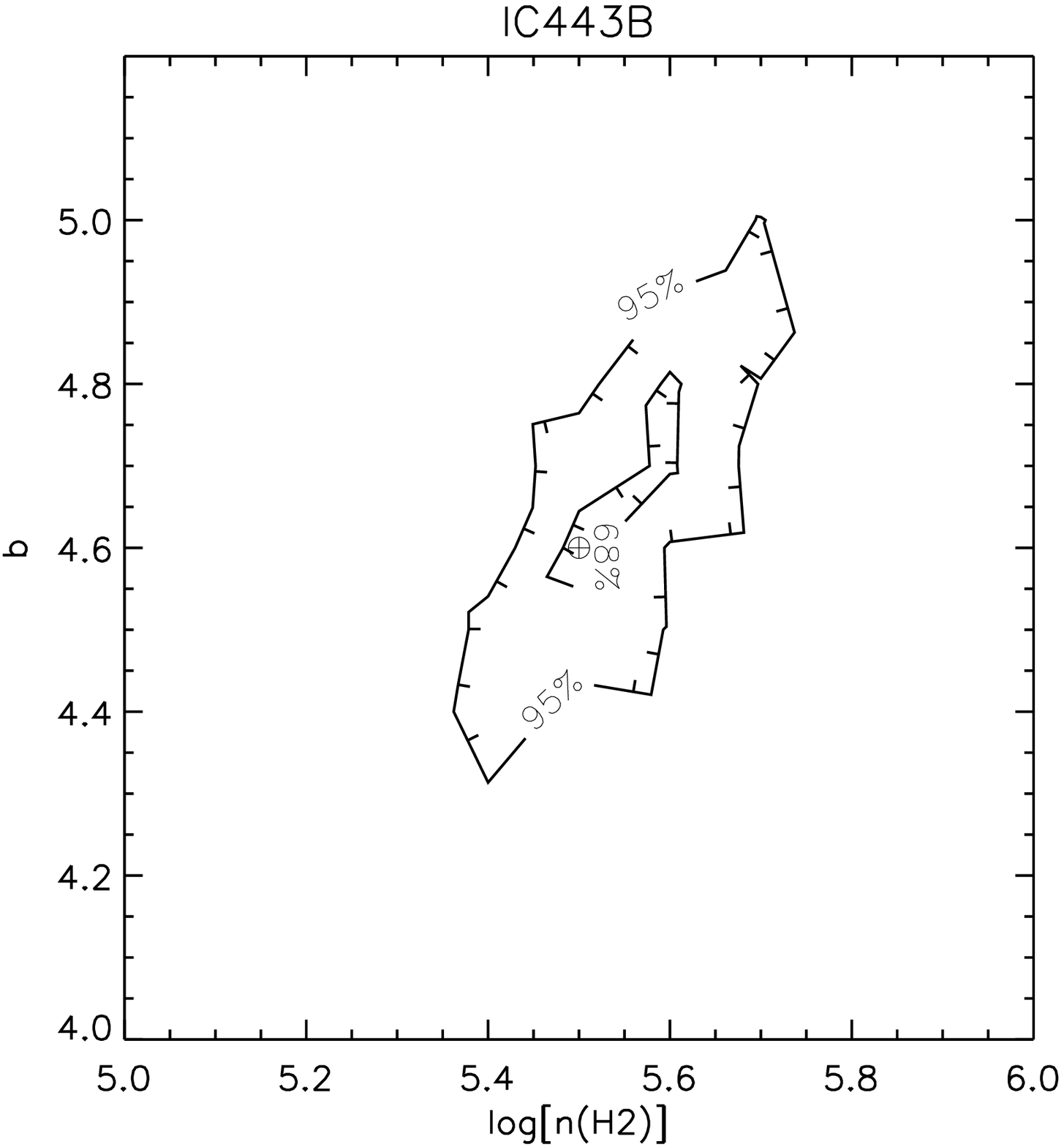}
\includegraphics[scale=0.45]{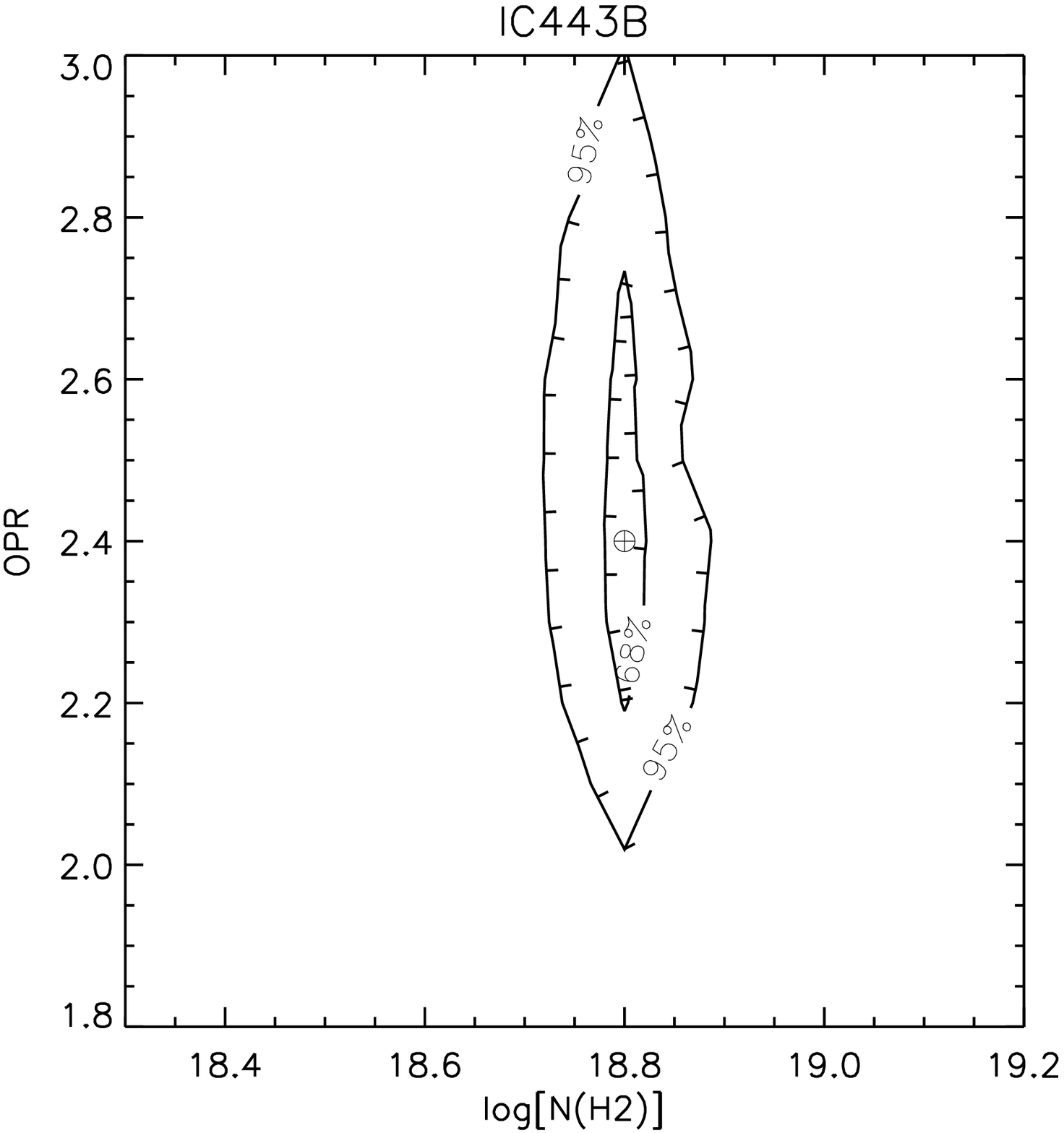}
\includegraphics[scale=0.45]{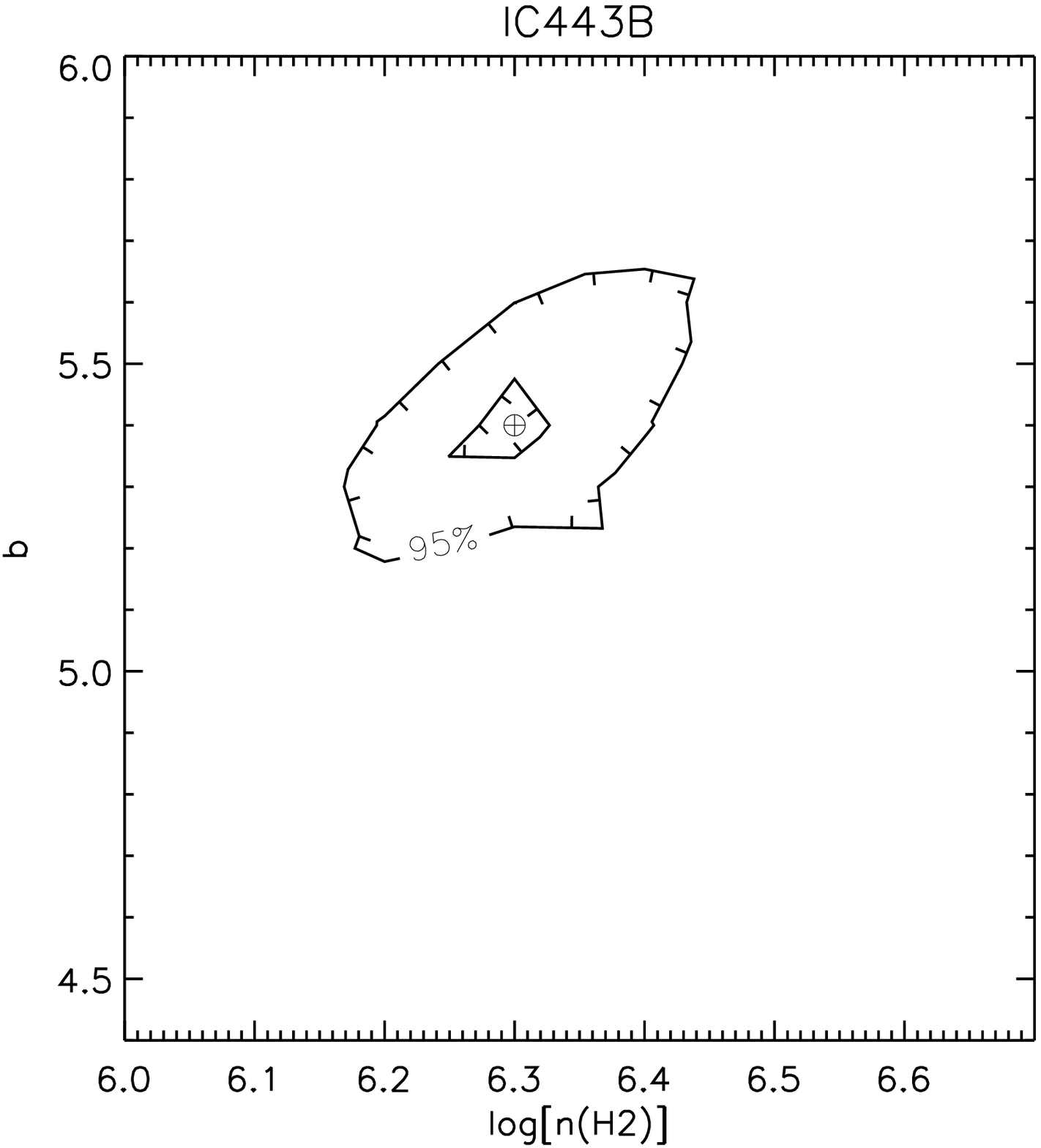}
\includegraphics[scale=0.45]{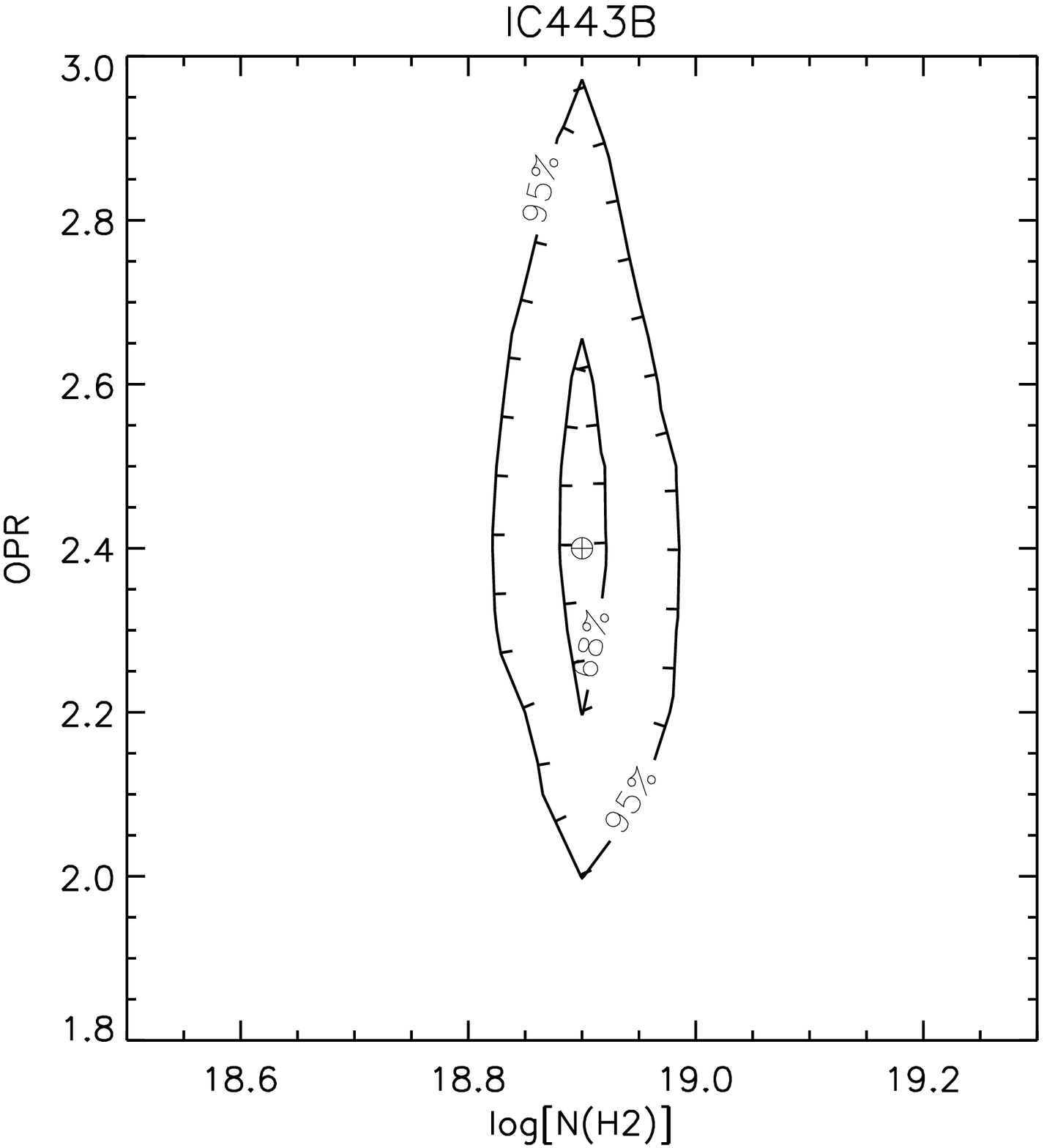}
}
\caption{The contour of $\chi^2$ in the plane of model parameters for IC 443B: X$_H=-1.7$ (\emph{top}) and X$_H=-2.5$ (\emph{bottom}) (cf. Table \ref{tbl-mfit-spec} and Fig.~\ref{fig-mfit-spec}). The 68\% and 95\% confidence levels are outlined. The tick marks along the contours indicate the directions that $\chi^2$ values are decreasing. `$\oplus$' indicates the model parameter values whose $\chi^2$ is minimum; i.e., the best fit. \label{fig-mconf-spec}}
\end{figure}

In order to investigate the OPR of shocked \Htwo{} gas in IC 443B using all possible observational constraints, we additionally performed spectral fitting to the observed \akari{}-IRC spectrum of IC 443B.
The model spectrum is made by convolving the modeled intensity with a Gaussian kernel whose full-width-at-half-maximum is 0.03 \um.
The extinction effect is included using $A_V=13.5$ \citep{Shinn(2011)ApJ_732_124} and the extinction curve of ``Milky Way, $R_V=3.1$'' \citep{Weingartner(2001)ApJ_548_296,Draine(2003)ARA&A_41_241}.
We used a synthetic continuum that is made by median-smoothing the observed spectrum with a 0.29 \um{} box kernel, and the continuum is fixed during the fitting.
The calibration systematic error of 10 \% (cf.~section \ref{obs-red}) is quadratically included in the error of spectrum.
The wavelength shift is included as a free parameter, and found to be $\sim+0.008$ \um{} for the best fit.
The fitting results are shown in Figure \ref{fig-mfit-spec}.
We fitted the spectrum fixing $X_H=-1.7$ and $X_H=-2.5$ as in the level population fitting (Fig.~\ref{fig-mfit}).
The best fit model parameters and the $\chi^2$ contours are shown in Table \ref{tbl-mfit-spec} and Figure \ref{fig-mconf-spec}.

The OPR is found to be $2.4^{+0.3}_{-0.2}$, which is a little higher than the ones from the level population fitting (Table \ref{tbl-mfit}).
However, the OPR is still less than 3.0 with a 95\% significance limit (Fig.~\ref{fig-mconf-spec}).
The index $b$ also increases a little, while \NHtwo{} and \nHtwo{} are almost the same.
The difference between two kinds of fitting is likely induced by the constraints from the weak and blended \Htwo{} emission lines, which are ignored in the level population fitting (Fig.~\ref{fig-mfit}).
Such contributions can be seen in Table \ref{tbl-mspec}; for example, the three lines  around 4.41 \um, $\upsilon=0-0$ S(10), $\upsilon=1-1$ S(11), and $\upsilon=2-1$ O(8).

For the HB 21 data, we could not get the range of OPR, because there are only a few emission lines strong enough to constraint the model parameters.
We estimate the OPRs of HB 21N and HB 21S with a single $T_{ex}$ assumption using the 4.2, 4.4, 4.7 \um{} emission lines in section \ref{ana-res-pop}.
They are found to be $1.7-1.8$.
If the contribution of $\upsilon=1-1$ S(11) 4.42 \um{} to the 4.4 \um{} line feature and the contribution of $\upsilon=1-0$ O(8) 4.16 \um{} to the 4.2 \um{} line feature are respectively around 13\% and 10\% as in IC 443B (cf.~Table \ref{tbl-mspec}), then the OPRs increase to $1.8-2.0$.
However, this estimation is still tentative.
More near-infrared \Htwo{} emission line data are required to assess the OPR of shocked \Htwo{} gas in HB 21.

\section{Discussion \label{discu}}
\subsection{Origin of the Non-equilibrium Ortho-to-Para Ratio Obtained from \EvJ{} $\ga7000$ K\label{dis-ori}}
We obtained the results suggesting shocked \Htwo{} gas of non-equilibrium OPRs in IC 443B (OPR$=2.4^{+0.3}_{-0.2}$), HB 21N and HB 21S (OPR$=1.8-2.0$, cf.~section \ref{ana-res-plmod-spec}).
%We obtained non-equilibrium OPRs ($1.6-2.2$) from the level populations of shocked \Htwo{} gas ($1000\le T\le 4000$ K) in IC 443B, HB 21N, and HB 21S (Table \ref{tbl-mfit}).
First, we examine the possibility that this non-equilibrium OPR originates from dissociative J-shocks, based on the results of \cite{Shinn(2011)ApJ_732_124}.
They studied shocked \Htwo{} gas in IC 443 (IC 443C and IC 443G), and found that the levels we here study ($7000\la$ \EvJ{} $\la20000$ K) are likely excited by dissociative J-shocks.
Dissociative J-shocks mainly excite those levels at the \Htwo{} reformation zone \citep[cf.][]{Flower(2003)MNRAS_341_70} through two main excitation mechanisms: the \Htwo{} formation pumping and the collisions with ambient gas \citep{Hollenbach(1989)ApJ_342_306}.
Therefore, the non-equilibrium OPR most likely originate from the \Htwo{} reformation zone, if dissociative J-shocks are responsible for the level excitation of 7000 K $\la$ \EvJ{} $\la$ 20000 K.

When the levels of 7000 K $\la$ \EvJ{} $\la$ 20000 K is dominantly populated, the temperature is a few hundred kelvin \citep{Flower(2003)MNRAS_341_70}.
This temperature is too low to render the ortho-para interconversion.
For example, the number of para-to-ortho conversion is about 0.6 for the typical J-shock seen in \cite{Flower(2003)MNRAS_341_70}.
We use the rate coefficient $k_{po}$ of $8 \times 10^{-11}$ exp($-3900/T$) cm$^3$ s$^{-1}$ for \astH{I} \citep{Schofield(1967)P&SS_15_643}, \nHone{}$=10^6$ \ncm{}, $t=100$ yr, and $T=300$ K.
Therefore, the OPR upon reformation behind dissociative J-shocks would remain unchanged until the gas cools down as low as the preshock gas.
This means that the non-equilibrium OPR should be acquired when \Htwo{} is reformed, if dissociative J-shocks are responsible.

\Htwo{} reforms on grain surfaces or through the gas phase reaction (\citealt{vandeHulst(1948)HarMo_7_73,McDowell(1961)Obs_81_240}; \citealt{Field(1966)ARA&A_4_207}, \citealt{Shull(1982)ARA&A_20_163}, and references therein).
Within the \Htwo{} reformation zone of dissociative J-shocks, the grain surface reformation is dominant, because the gas phase reformation is inefficient when the ionization fraction of shocked gas is below 2\% \citep{Neufeld(1989)ApJ_340_869}.
The OPR of reformed \Htwo{} was usually \emph{assumed} to be the equilibrium OPR in dissociative J-shock models \citep{Hollenbach(1989)ApJ_342_306,Flower(2003)MNRAS_341_70}, following the studies on the level population of newly formed \Htwo{} \citep{Black(1976)ApJ_203_132,Bieniek(1979)ApJ_228_635,Bieniek(1980)JPhB_13_4405,Black(1987)ApJ_322_412,Flower(2003)MNRAS_341_70}.
Thus, we need to more scrutinize previous theoretical and experimental studies relevant to the OPR of newly formed \Htwo.

As far as we are aware, there are two studies that performed a quantum calculation for the \Htwo{} formation upon grains and mentioned the OPR \citep{Meijer(2001)JPCA_105_2173,Morisset(2005)JChPh_122_194702}.
Among them, the results of \cite{Morisset(2005)JChPh_122_194702} do not exclude the possibility that the newly-formed \Htwo{} has an OPR $=1.8-2.4$ (cf.~section \ref{ana-res-plmod-spec}).
\cite{Meijer(2001)JPCA_105_2173} and \cite{Morisset(2005)JChPh_122_194702} respectively studied the Eley-Rideal (ER) and Langmuir-Hinshelwood (LH) formation mechanisms\footnote{See \cite{Tielens(2005)book} for the ER and LH mechanisms.}, among which the LH mechanism is more relevant to the interstellar environments \citep{Pirronello(2000)inproc}.
The level population upon formation is biased within $\upsilon\ga5$ for the LH mechanism, while it is scattered over all the considered vibrational levels ($\upsilon=0-6$) for the ER mechanism.
They both calculated the OPR, and it is fluctuating within $\sim1.5-6.0$ over the collision energy of $4-50$ meV in the LH mechanism, while it is close to 3 over the initial translational energy of $0-200$ meV in the ER mechanism.
We think that the OPR$=1.8-2.4$ can be acquired through the LH formation mechanism, if the collision energy of H atoms on grain surfaces has an appropriate distribution.
Note that the biased level population ($\upsilon\ga5$) of the LH mechanism is redistributed over all levels through collisions with ambient postshock gas (i.e.~ortho-to-ortho and para-to-para transitions).

We found two experimental works that studied the OPR of newly-formed \Htwo{} on grains \citep{Perry(2003)Ap&SS_285_769,Creighan(2006)JChPh_124_114701}.
However, their results cannot tell if the newly-formed \Htwo{} can have an OPR $=1.8-2.4$ or not, because of the experimental contaminations.
They measured the level population of newly-formed \Htwo{} on a graphite surface, called ``highly oriented pyrolytic graphite,'' at a base pressure of $\sim10^{-10}$ Torr ($\sim10^9$ cm$^{-3}$ K) much higher than the typical pressure of interstellar clouds, $10^4-10^6$ cm$^{-3}$ K \citep{Bergin(2007)ARA&A_45_339}. 
The obtained population is close to the LTE.
However, it was known that the measured \Htwo{} population had been contaminated by the undissociated \Htwo{} that originates from the H atom source \citep{Islam(2010)ApJ_725_1111}.

From the above discussion, we show that the non-equilibrium OPRs of $1.8-2.4$ can originate from dissociative J-shocks.
%From the above discussion, we show that the non-equilibrium OPRs of IC 443B, HB 21N, and HB 21S (Table \ref{tbl-mfit}) can originate from dissociative J-shocks.
We here note that the dissociative J-shock interpretation is not incompatible with the observed line ratios between \Htwo{} and \astH{I}.
We compare the intensities of Br$\alpha$ 4.05 \um{} to those of $\upsilon=0-0$ S(11) 4.18 \um.
Table \ref{tbl-bra} shows the expected and observed intensity of Br$\alpha$.
The expected Br$\alpha$ intensity is calculated from the observed $\upsilon=0-0$ S(11) intensity \citep[][Table \ref{tbl-result}]{Shinn(2011)ApJ_732_124}, employing the dissociative J-shock model of \cite{Hollenbach(1989)ApJ_342_306}.
We ignore the effect of extinction, because the wavelengths of Br$\alpha$ and $\upsilon=0-0$ S(11) are similar.
The adopted shock model parameters are $n_0=10^3$ \ncm, $v_s=40-80$ \kms.
Note that the dissociative J-shocks are likely propagating into interclump media of $<10^3$ \ncm, according to \cite{Shinn(2011)ApJ_732_124}.
As seen from Table \ref{tbl-bra}, the observed intensities better agree with the model expectations, when the shock velocity is slower.
The intensities of $\upsilon=0-0$ S(11) are also compatible with the model values, considering the possibilities of the multiple shock surfaces, the shock surface viewed edge-on, the not-fully-covered shock surface because of unresolved substructures, etc.

Now we examine other possibilities than the dissociative J-shocks.
The shocked \Htwo{} gas of non-equilibrium OPR can be made by C-shocks or partially-dissociative J-shocks \citep{Timmermann(1998)ApJ_498_246,Wilgenbus(2000)A&A_356_1010}.
However, these shocks are incapable of emanating the ionic lines observed in the shocked gas.
When \cite{Shinn(2011)ApJ_732_124} interpreted the level population of shocked \Htwo{} gas in IC 443C and IC 443G (0 $\la$ \EvJ{} $\la$ 25000 K), they concluded that a combination of C-shocks and dissociative J-shocks is needed.
Under this interpretation, the dissociative J-shocks are mandatory to explain the observed ionic emission lines, like [\astNe{II}] 12.8 \um{} and [\astC{II}] 158 \um.
Such lines were also observed at IC 443B \citep{Noriega-Crespo(2009)inproc}.
In HB 21, no detection of such ionic lines has been reported.
However, the level population shape of shocked \Htwo{} gas over 0 K $\la$ \EvJ{} $\la$ 20000 K is similar to that of IC 443C and IC 443G \citep[Fig.~\ref{fig-pop},][]{Shinn(2009)ApJ_693_1883,Shinn(2010)AdSpR_45_445,Shinn(2011)ApJ_732_124}---the power-law thermal admixture model with two density components is required to reproduce the level population.
As in IC 443, dissociative J-shocks may be needed to explain the level population of shocked \Htwo{} gas in HB 21.
We thus think that the non-equilibrium OPR is unlikely to originate from either C-shocks or partially-dissociative J-shocks.

\subsection{Ortho-to-Para Ratios of Shocked \Htwo{} Gas in IC 443 and HB 21}
The OPR obtained from the levels of 7000 K $\la$ \EvJ{} $\la$ 20000 K is probably in non-equilibrium for IC 443B (section \ref{ana-res-plmod-spec}), while it is in equilibrium for IC 443C and IC 443G \citep{Shinn(2011)ApJ_732_124}.
%The OPR obtained from the levels of 7000 K $\la$ \EvJ{} $\la$ 20000 K is the non-equilibrium OPR for IC 443B (Table \ref{tbl-mfit}), while it is the equilibrium OPR for IC 443C and IC 443G \citep{Shinn(2011)ApJ_732_124}.
If we accept the conclusion from section \ref{dis-ori}, these OPRs are determined by the collision energy distribution of H atoms on grain surfaces.
Therefore, the resultant OPRs means that the collision energy distribution is similar each other at IC 443C and IC 443G, while it is different for IC 443B.
It is uncertain how the collision energy distribution is determined.
The property of grain surface and the environment where grains and \astH{I} gas are embedded may be included in the determining factors.
We here note that IC 443B is located at the vertex of the ``W-shaped'' \Htwo{} feature, while IC 443C and IC 443G are not \citep[cf.~Fig.~1 of][]{Shinn(2011)ApJ_732_124}.
This implies that the density at IC 443B may be higher than the others \citep[cf.][]{Shinn(2011)ApJ_732_124}.

The OPRs of HB 21N and HB 21S are similar each other (Table \ref{tbl-mfit}).
This indicates the collision energy distribution is similar at both regions, if we accept the conclusion from section \ref{dis-ori}.
The OPRs of HB 21 are lower than those of IC 443.
HB 21 shows a lower \Htwo{} column density than IC 443 \citep{Neufeld(2007)ApJ_664_890,Shinn(2009)ApJ_693_1883,Shinn(2010)AdSpR_45_445,Shinn(2011)ApJ_732_124}.
This implies a lower density for HB 21, if a similar extension of shocked \Htwo{} gas along the line of sight is assumed for IC 443 and HB 21.

If our inference for the preshock density of IC 443 and HB 21 is the case, the preshock density and the obtained OPR show neither a correlation nor an anti-correlation.
They are anti-correlated for IC 443B, IC 443C, and IC 443G, while they are correlated for IC 443 and HB 21.
This indicates that the obtained OPRs are not simply determined by the preshock density.
More information on physical parameters is required to understand how the obtained OPRs are determined at high energy levels of 7000 K $\la$ \EvJ{} $\la$ 20000 K.
%More detailed shock model studies for diverse shock parameters would be helpful in understanding the obtained OPR $=1.6-2.2$ at high energy levels of 7000 K $\la$ \EvJ{} $\la$ 20000 K.

\section{Conclusions \label{concl}}
We present the near-infrared ($2.5-5.0$ \um) spectral observations of shocked \Htwo{} gas in the two SNRs IC 443 and HB 21.
The observed regions are IC 443B, HB 21N, and HB 21S.
At the energy range 7000 K $\la$ \EvJ{} $\la$ 20000 K, IC 443B shows an OPR of $2.4^{+0.3}_{-0.2}$ ($<3$ in 95\% confidence limit), which suggests the existence of non-equilibrium OPR. 
HB 21N and HB 21S also show indications of non-equilibrium OPRs in the range of $1.8-2.0$.
These may be the first objects observed with non-equilibrium OPR at such a high \EvJ{} range.
%IC 443B shows \Htwo{} level populations with probable non-equilibrium OPRs ($=2.4^{+0.3}_{-0.2}$, $<3.0$ in 95\% confidence limit) at 7000 K $\la$ \EvJ{} $\la$ 20000 K, while HB 21N and HB 21S tentatively show signs of non-equilibrium OPRs ($1.8-2.0$).
%This is the first time that reports a probable non-equilibrium OPR at such a high \EvJ{} range.
%All the observed regions (IC 443B, HB 21N, and HB 21S) show \Htwo{} level populations with non-equilibrium OPRs ($=1.6-2.2$) with $1000\le T \le 4000$ K at 7000 K $\la$ \EvJ{} $\la$ 20000 K.
The level populations are well described by the power-law thermal admixture model ($1000\le T \le 4000$ K) with a single OPR, where the infinitesimal column density is proportional to the inverse power of the temperature.

We conclude that the obtained non-equilibrium OPR probably originates from the \Htwo{} reformation through dissociative J-shocks, based on several arguments.
\cite{Shinn(2011)ApJ_732_124} found that the \Htwo{} levels of $7000\la$ \EvJ{} $\la20000$ K are likely excited by dissociative J-shocks, where \Htwo{} is dissociated and then reformed.
Previous quantum calculations do not exclude the possibility that the reformed \Htwo{} has an OPR of $1.8-2.4$, when formed with an appropriate collision energy of H atoms on grain surfaces.
The observed line ratio between \Htwo{} $\upsilon=0-0$ S(11) and Br$\alpha$ is also compatible with the dissociative J-shock interpretation.

According to our conclusion, the difference in the obtained OPRs is caused by the different collision energy of H atoms on grain surfaces.
It is uncertain how the collision energy is determined, but the property of grain surface and the environment where grains and \astH{I} gas are embedded may be included in the determining factors.
We check if any simple relation is held between the obtained OPR and preshock density for IC 443 and HB 21, but it turns out unlikely.

More observational studies are required to confirm the non-equilibrium OPR; to validate our explanations for the obtained non-equilibrium OPR; and to grasp how a non-equilibrium OPR of shocked \Htwo{} gas is determined at the energy levels of 7000 K $\la$ \EvJ{} $\la$ 20000 K.
Our study suggests that dissociative J-shocks can make shocked \Htwo{} gas with a non-equilibrium OPR.

\acknowledgments
This work is based on observations with \akari, a JAXA project with the participation of ESA. 
The authors thank all the members of the \akari{} project.
The authors also appreciate the referee's comments, which improve the original manuscript.
J. H. S. is grateful to Bon-Chul Koo and Michael Burton for the discussion that clarified the ortho-to-para ratio and related modeling, and grateful to Ji Yeon Seok for the useful discussion on the data reduction.
This research has made use of SAOImage DS9, developed by Smithsonian Astrophysical Observatory \citep{Joye(2003)inproc}.

%%%%%%%%%% FIGURES
\clearpage

\clearpage

\clearpage

\clearpage

\clearpage

\clearpage

\clearpage

\clearpage

\clearpage

%%%%%%%%%% TABLES
\clearpage
\begin{deluxetable}{cccc}
\tablewidth{0pt}
\tablecaption{Summary of the \akari{} IRC Observations \label{tbl-obs}}
\tablehead{
\colhead{Region} &  \colhead{Pointing Position} & \colhead{Observation ID} & \colhead{AOT\tablenotemark{a}} \\
%& & \colhead{coverage\tablenotemark{a}} & \colhead{Resolution ($\Gamma$)}\\
& \colhead{(RA, Dec; J2000)} & }

\startdata
HB 21N		& (20:47:26.891,+51:15:04.80)	& 1420797.[1-4], 1420798.1 & IRCZ4\\
HB 21N BG	& (20:47:32.237,+51:16:34.86)	& 1420799.[1-2] & IRCZ4\\
HB 21S		& (20:46:10.398,+49:58:15.20)	& 1420800.[1-4], 1420801.1 & IRCZ4\\
HB 21S BG	& (20:46:04.220,+49:58:13.61)	& 1420802.[1-2] & IRCZ4\\
\enddata
\tablenotetext{a}{Astronomical Observation Template. It is a pre-defined observation sequence. See the IRC Data User Manual for Post-Helium Mission. \citep{Onaka(2009)mana}.}
\tablecomments{The observation summary for the IC 443B data is in \cite{Shinn(2011)ApJ_732_124}.}
\end{deluxetable}

\begin{deluxetable}{lcrrrr}
%\tabletypesize{\tiny}
%\tabletypesize{\scriptsize}
%\tabletypesize{\footnotesize}
%\tabletypesize{\small}
\tablewidth{0pt}
\tablecaption{Observed \Htwo{} Emission Lines \label{tbl-result}}
\tablehead{
\colhead{Transition} & \colhead{Wavelength} & \multicolumn{2}{c}{Observed Intensity} \\
&($\mu$m)	&\multicolumn{2}{c}{($10^{-6}$ erg s$^{-1}$ cm$^{-2}$ sr$^{-1}$)}\\
\cline{3-4}
&	&\colhead{HB 21N} &\colhead{HB 21S}
}
\startdata

$\upsilon=$ 1-0 O(3)&   2.80&       9.1$\pm$2.0&      11.6$\pm$2.0\\
$\upsilon=$ 2-1 O(3)&   2.97&            $<$4.1&           \nodata\\
$\upsilon=$ 1-0 O(4)&   3.00&            $<$5.7&           \nodata\\
$\upsilon=$ 1-0 O(5)&   3.23&            $<$6.2&       8.3$\pm$1.7\\
$\upsilon=$ 1-0 O(7)&   3.81&            $<$3.8&           \nodata\\
$\upsilon=$ 0-0 S(13)&   3.85&            $<$4.9&           \nodata\\
$\upsilon=$ 0-0 S(11)&   4.18&       8.4$\pm$1.4&       9.0$\pm$1.3\\
$\upsilon=$ 0-0 S(10)&   4.41&       7.5$\pm$1.4&       8.7$\pm$1.3\\
$\upsilon=$ 0-0 S(9)&   4.69&      19.7$\pm$2.4&      21.6$\pm$2.5\\

\enddata
%\tablenotetext{a}{in units of \luerg}
%\tablenotetext{a}{Extinction-corrected intensity with $N$(H)=$3.5\times10^{21}$ \Ncm.}
%\tablecomments{The three regions are indicated in Figure \ref{fig-rgb}. The upper limits are 90\% upper confidence limits.}
\tablecomments{For those lines whose significance is lower than 3.0, the intensities are expressed with 90\% upper confidence limits. The error includes both statistical and systematic components. See text for detail.}
\end{deluxetable}

\begin{deluxetable}{lrrrrr}
%\tabletypesize{\tiny}
%\tabletypesize{\scriptsize}
%\tabletypesize{\footnotesize}
%\tabletypesize{\small}
\tablewidth{0pt}
\tablecaption{Extinction Corrected \Htwo{} Column Density \label{tbl-h2col}}
\tablehead{
\colhead{Level} & \colhead{Level Energy} & \multicolumn{2}{c}{log N(\Htwo; $\upsilon,J$)}\\
($\upsilon,J$)  & \colhead{(K)} &\multicolumn{2}{c}{(cm$^{-2}$)}\\
\cline{3-4}
&   &\colhead{HB 21N} &\colhead{HB 21S}
}
\startdata

(0,11)&    10261.&          15.10$\pm$0.05&          15.14$\pm$0.05\\
(0,12)&    11940.&          14.50$\pm$0.08&          14.56$\pm$0.07\\
(0,13)&    13703.&          14.39$\pm$0.07&          14.42$\pm$0.06\\
(0,15)&    17443.&                $<$13.90&                 \nodata\\
(1, 1)&     6149.&          14.63$\pm$0.10&          14.74$\pm$0.07\\
(1, 2)&     6471.&                $<$14.62&                 \nodata\\
(1, 3)&     6951.&                $<$14.82&          14.95$\pm$0.09\\
(1, 5)&     8365.&                $<$14.97&                 \nodata\\
(2, 1)&    11789.&                $<$14.13&                 \nodata\\

\enddata
\tablecomments{The extinctions were corrected, using \defNH$=3.5\times10^{21}$ \Ncm{} \citep{Lee(2001)inproc} and the extinction curve of ``Milky Way, $R_V=3.1$'' \citep{Weingartner(2001)ApJ_548_296,Draine(2003)ARA&A_41_241}.}
%\tablenotetext{b}{The population cannot be not determined because of the probable line blending with nearby lines. cf.~Table \ref{tbl-result}.}
%\tablecomments{The three regions are indicated in Figure \ref{fig-rgb}. The upper limits are 90\% upper confidence limits.}
\end{deluxetable}

\clearpage
\begin{deluxetable}{ccccccc}
%\tabletypesize{\tiny}
%\tabletypesize{\scriptsize}
%\tabletypesize{\footnotesize}
%\tabletypesize{\small}
\tablewidth{0pt}
\tablecaption{Level Population Fitting Results for the Model Parameters \label{tbl-mfit}}
\tablehead{
%&\multicolumn{2}{c}{component 1} & &\multicolumn{2}{c}{component 2} \\
%\cline{2-3} \cline{5-6}
\colhead{Region} & \colhead{log[$N$(H$_2$)]} & \colhead{log[$n$(H$_2$)]} & \colhead{$b$} & \colhead{X$_H$\tablenotemark{a}} & \colhead{OPR} & \colhead{${\chi}^2_{\nu}$} \\
 & \colhead{(cm$^{-2}$)} & \colhead{(cm$^{-3}$)} &  & \colhead{$\left(\equiv\mathrm{log}\left[\frac{n(\mathrm{H I})}{n(\mathrm{H}_2)}\right]\right)$} & & \colhead{$(\equiv\chi^2$/d.o.f)}
}
\startdata

IC443B&$18.6_{-0.1}^{+0.1}$&$5.8_{-0.2}^{+0.4}$&$3.9_{-0.4}^{+0.4}$&$-1.7$&$2.1_{-0.2}^{+0.2}$&1.8 (=7.2/4.0)
\\
IC443B&$18.8_{-0.2}^{+0.1}$&$6.5_{-0.2}^{+0.3}$&$4.5_{-0.5}^{+0.5}$&$-2.5$&$2.2_{-0.3}^{+0.2}$&1.4 (=5.5/4.0)
\\
HB21N&$17.4$&$5.5$&$2.6$&$-2.0$&$1.8$&\nodata
\\
HB21S&$17.7$&$5.5$&$3.3$&$-2.0$&$1.6$&\nodata
\\

\enddata
\tablenotetext{a}{The parameter $X_H$ is fixed as shown. See section \ref{ana-res-plmod} for more detail.}
%\tablenotetext{b}{The population is not determined because of the line blending. cf.~Table \ref{tbl-result}.}
\tablecomments{The confidence limits are given with a 68\% significance (cf.~Fig.~\ref{fig-mconf}). \NHtwo{} means $N$(\Htwo; $T>1000$ K). See section \ref{ana-res-plmod} for the detailed description about the parameters.}
\end{deluxetable}

\clearpage
\begin{deluxetable}{ccccccc}
%\tabletypesize{\tiny}
%\tabletypesize{\scriptsize}
%\tabletypesize{\footnotesize}
%\tabletypesize{\small}
\tablewidth{0pt}
\tablecaption{Spectral Fitting Results for the Model Parameters \label{tbl-mfit-spec}}
\tablehead{
%&\multicolumn{2}{c}{component 1} & &\multicolumn{2}{c}{component 2} \\
%\cline{2-3} \cline{5-6}
\colhead{Region} & \colhead{log[$N$(H$_2$)]} & \colhead{log[$n$(H$_2$)]} & \colhead{$b$} & \colhead{X$_H$\tablenotemark{a}} & \colhead{OPR} & \colhead{${\chi}^2_{\nu}$} \\
 & \colhead{(cm$^{-2}$)} & \colhead{(cm$^{-3}$)} &  & \colhead{$\left(\equiv\mathrm{log}\left[\frac{n(\mathrm{H I})}{n(\mathrm{H}_2)}\right]\right)$} & & \colhead{$(\equiv\chi^2$/d.o.f)}
}
\startdata

IC443B&$18.8_{-0.02}^{+0.02}$&$5.5_{-0.04}^{+0.11}$&$4.6_{-0.05}^{+0.21}$&$-1.7$&$2.4_{-0.2}^{+0.3}$&1.93 (=408.7/212.0)
\\
IC443B&$18.9_{-0.02}^{+0.02}$&$6.3_{-0.05}^{+0.03}$&$5.4_{-0.05}^{+0.08}$&$-2.5$&$2.4_{-0.2}^{+0.3}$&1.89 (=401.4/212.0)
\\

\enddata
\tablenotetext{a}{The parameter $X_H$ is fixed as shown. See section \ref{ana-res-plmod} for more detail.}
%\tablenotetext{b}{The population is not determined because of the line blending. cf.~Table \ref{tbl-result}.}
\tablecomments{The confidence limits are given with a 68\% significance (cf.~Fig.~\ref{fig-mconf-spec}). \NHtwo{} means $N$(\Htwo; $T>1000$ K). See section \ref{ana-res-plmod} for the detailed description about the parameters.}
\end{deluxetable}

\clearpage
\begin{deluxetable}{lcrr}
%\tabletypesize{\tiny}
%\tabletypesize{\scriptsize}
%\tabletypesize{\footnotesize}
%\tabletypesize{\small}
\tablewidth{0pt}
\tablecaption{\Htwo{} Emission Lines from the Best Fit Models of the IC 443B data \label{tbl-mspec}}
\tablehead{
\colhead{Transition} & \colhead{Wavelength} & \multicolumn{2}{c}{Modeled Intensity} \\
&($\mu$m)   &\multicolumn{2}{c}{($10^{-6}$ erg s$^{-1}$ cm$^{-2}$ sr$^{-1}$)}\\
\cline{3-4}
&   &\colhead{$X_H=-1.7$} &\colhead{$X_H=-2.5$}
}
\startdata

$\upsilon=$2-1 Q(9) &      2.72 &       1.7 &       1.3\\
$\upsilon=$1-0 Q(13) &      2.73 &       0.8 &       0.7\\
$\upsilon=$2-1 O(2) &      2.79 &       2.1 &       1.7\\
$\upsilon=$1-0 O(3) &      2.80 &      66.9 &      69.0\\
$\upsilon=$2-1 O(3) &      2.97 &       6.9 &       5.8\\
$\upsilon=$1-0 O(4) &      3.00 &      28.5 &      29.2\\
$\upsilon=$2-1 O(4) &      3.19 &       3.1 &       2.6\\
$\upsilon=$1-0 O(5) &      3.23 &      53.9 &      53.8\\
$\upsilon=$2-1 O(5) &      3.44 &       6.2 &       5.1\\
$\upsilon=$1-0 O(6) &      3.50 &      15.5 &      15.3\\
$\upsilon=$2-1 O(6) &      3.72 &       1.9 &       1.6\\
$\upsilon=$1-0 O(7) &      3.81 &      21.6 &      20.5\\
$\upsilon=$0-0 S(13) &      3.85 &      15.2 &      16.0\\
$\upsilon=$0-0 S(12) &      4.00 &      11.2 &      11.4\\
$\upsilon=$2-1 O(7) &      4.05 &       2.9 &       2.3\\
$\upsilon=$1-0 O(8) &      4.16 &       4.8 &       4.5\\
$\upsilon=$0-0 S(11) &      4.18 &      44.6 &      45.0\\
$\upsilon=$0-0 S(10) &      4.41 &      30.5 &      30.9\\
$\upsilon=$1-1 S(11) &      4.42 &       4.6 &       4.1\\
$\upsilon=$2-1 O(8) &      4.44 &       0.7 &       0.6\\
$\upsilon=$1-0 O(9) &      4.58 &       5.4 &       4.8\\
$\upsilon=$0-0 S(9) &      4.69 &     111.0 &     117.6\\

\enddata
%\tablenotetext{a}{in units of \luerg}
%\tablenotetext{a}{Extinction-corrected intensity with $N$(H)=$3.5\times10^{21}$ \Ncm.}
%\tablecomments{The three regions are indicated in Figure \ref{fig-rgb}. The upper limits are 90\% upper confidence limits.}
\tablecomments{The best fit models are from the spectral fitting (Figure \ref{fig-mfit-spec}) and the extinction effect is applied.}
\end{deluxetable}

\clearpage
\begin{deluxetable}{ccc}
%\tabletypesize{\tiny}
%\tabletypesize{\scriptsize}
%\tabletypesize{\footnotesize}
%\tabletypesize{\small}
\tablewidth{0pt}
\tablecaption{Comparison between the Expected and Observed Intensities of Br$\alpha$ \label{tbl-bra}}
\tablehead{
%&\multicolumn{2}{c}{component 1} & &\multicolumn{2}{c}{component 2} \\
%\cline{2-3} \cline{5-6}
\colhead{Region} & \colhead{Expected\tablenotemark{a}} & \colhead{Observed\tablenotemark{b}} \\
&\multicolumn{2}{c}{(erg s$^{-1}$ cm$^{-2}$ sr$^{-1}$)}
}
\startdata

    IC443B &           (9.3e-07,\,5.6e-05) &  $<$1.7e-05\\
    IC443C &           (3.1e-06,\,1.9e-04) &  $<$1.8e-05\\
    IC443G &           (1.8e-06,\,1.1e-04) &  $<$2.0e-05\\
     HB21N &           (1.3e-07,\,7.8e-06) &  $<$4.8e-06\\
     HB21S &           (1.4e-07,\,8.4e-06) &  $<$4.7e-06\\

\enddata
\tablenotetext{a}{The expected intensity range of Br$\alpha$ is calculated from the observed intensity of $\upsilon=0-0$ S(11) \citep[][Table \ref{tbl-result}]{Shinn(2011)ApJ_732_124}, employing the dissociative J-shock model of \cite{Hollenbach(1989)ApJ_342_306}. The values are calculated for $n_0=10^3$ \ncm{} and $v_s=40-80$ \kms.}
\tablenotetext{b}{99 \% upper confidence limit.}
%\tablecomments{The confidence limits are given with a 68\% significance (cf.~Fig.~\ref{fig-mconf}). \NHtwo{} means $N$(\Htwo; $T>1000$ K). See section \ref{ana-res-plmod} for the detailed description about the parameters.}
\end{deluxetable}

\end{document}